%% file: main.tex
\PassOptionsToPackage{hyphens}{url}
\documentclass[sigconf,screen,nonacm]{acmart}
\usepackage{algorithmic}
\usepackage{textcomp}
\usepackage{tikz}
\usetikzlibrary{positioning,arrows.meta,fit,backgrounds,calc}
\usepackage{stfloats}
\usepackage{multirow}
\usepackage{lipsum}
\usepackage{subcaption}

\providecommand{\circled}[1]{%
  \tikz[baseline=(c.base)]\node[draw, circle, inner sep=0.6pt,
    line width=0.4pt, font=\footnotesize] (c) {#1};}

\newcounter{takeawaycount}
    
\newcommand{\takeaway}[1]{%
\refstepcounter{takeawaycount}
  \begin{center}
  \begin{tikzpicture}
    \node[draw=black!60, fill=black!20, rounded corners=2pt,
          inner sep=6pt, line width=0.4pt,
          text width=\dimexpr\linewidth-12.4pt\relax] {{\bf Observation \thetakeawaycount:}\emph{#1}};
  \end{tikzpicture}
  \end{center}}
\setcopyright{none}

\iftrue
\newcommand{\merkouriadis}[1]{{\footnotesize\color{purple}[merkouriadis: #1]}}
\newcommand{\rishabh}[1]{{\footnotesize\color{red}[Rishabh: #1]}}
\newcommand{\resolved}[1]{{\footnotesize\color{green}[Resolved: #1]}}
\newcommand{\TonyH}[1]{{\footnotesize\color{red}[TonyH: #1]}}
\newcommand{\JT}[1]{{\footnotesize\color{red}[JT: #1]}}
\newcommand{\HF}[1]{{\footnotesize\color{red}[HF: #1]}}

\newcommand{\YZ}[1]{{\footnotesize\color{red}[YZ: #1]}}
\newcommand{\CW}[1]{{\footnotesize\color{red}[CW: #1]}}
\newcommand{\MM}[1]{{\footnotesize\color{red}[MM: #1]}}
\else
\newcommand{\merkouriadis}[1]{}
\newcommand{\rishabh}[1]{}
\newcommand{\resolved}[1]{}
\newcommand{\TonyH}[1]{}
\newcommand{\JT}[1]{}
\newcommand{\HF}[1]{}
\newcommand{\FL}[1]{}
\newcommand{\YZ}[1]{}
\newcommand{\CW}[1]{}
\newcommand{\MM}[1]{}
\fi

\begin{document}

\title{Aggron: Aggressive Compression of Agentic KV Caches during Head-of-Line Stalls in the Request Queue}
\settopmatter{authorsperrow=4}

\author{Rishabh Ravi}
\affiliation{%
  \institution{University of Illinois Urbana-Champaign}
  \city{Urbana}
  \state{Illinois}
  \country{USA}}
\email{ravi15@illinois.edu}

\author{Tony Hong}
\affiliation{%
  \institution{University of Illinois Urbana-Champaign}
  \city{Urbana}
  \state{Illinois}
  \country{USA}}
\email{tonyhong@illinois.edu}

\author{Chen Wang}
\affiliation{%
  \institution{IBM Research}
  \city{Yorktown Heights}
  \state{New York}
  \country{USA}}
\email{chen.wang1@ibm.com}

\author{Yue Zhu}
\affiliation{%
  \institution{IBM Research}
  \city{Yorktown Heights}
  \state{New York}
  \country{USA}}
\email{yue.zhu@ibm.com}

\author{Mengmei Ye}
\affiliation{%
  \institution{IBM Research}
  \city{Yorktown Heights}
  \state{New York}
  \country{USA}}
\email{mye@ibm.com}

\author{Hubertus Franke}
\affiliation{%
  \institution{IBM Research}
  \city{Yorktown Heights}
  \state{New York}
  \country{USA}}
\email{frankeh@us.ibm.com}

\author{Fan Lai}
\affiliation{%
  \institution{University of Illinois Urbana-Champaign}
  \city{Urbana}
  \state{Illinois}
  \country{USA}}
\email{fanlai@illinois.edu}

\author{Josep Torrellas}
\affiliation{%
  \institution{University of Illinois Urbana-Champaign}
  \city{Urbana}
  \state{Illinois}
  \country{USA}}
\email{torella@illinois.edu}

\begin{abstract}
LLM agents run long, multi-turn programs whose contexts grow at every step, and serving them is limited by KV cache capacity. To alleviate this limitation, systems use context compression techniques, which
typically wait until the KV cache crosses a large threshold and then summarize large spans of history at once, either in the background while the agent progresses or on the critical path. These summarizations can be done using the same model running on the same engine, or using the same or a different model running
on extra hardware.
When running on the same engine, summarization competes with the requests being served for both compute and memory capacity.


To maximize the cost-effectiveness of summarization, this paper presents a new serving
system called {\em Aggron}. Aggron summarizes agentic requests at every turn and
selectively summarizes only certain parts of the KV cache. As a result, Aggron
retains accuracy and keeps summarization requests lightweight.
Moreover, Aggron launches summarization requests
from a separate {\em Summarization Queue} while the corresponding
agentic request 
whose KV cache 
we want to compress is
stalled in the main request queue due to head-of-line stalls.
The Summarization queue utilizes compute and memory capacity left over by the
head-of-line stalls in the main queue. Further, to maximize efficiency,
the server issues summarizations only when it predicts that they can complete before
the corresponding agentic request will finally execute.
In our evaluation, we compare Aggron to state-of-the-art asynchronous summarization.   
At high load, Aggron reduces P99
program latency by $27$--$31\%$, and raises throughput by
$1.34$--$1.40\times$.


\end{abstract}

 \maketitle


\section{Introduction}
\input{sections/1.intro}

\section{Background}
\input{sections/2.background}

\section{Opportunities for Compression of Agentic Requests}\label{sec:motivation}
\input{sections/3.PA}

\section{Aggron: Aggressive Compression}
\input{sections/4.aggron}

\section{Experimental Setup}\label{sec:setup}
\input{sections/5.setup}

\section{Evaluation}\label{sec:eval}
\input{sections/6.eval}

\section{Related Work}
\input{sections/7.related}

\section{Conclusions}

\input{sections/8.Conclusion}



\bibliographystyle{ACM-Reference-Format}

\bibliography{refs}

\end{document}

%% file: sections/1.intro.tex
Large Language Models (LLMs) are increasingly deployed as {agents} that
invoke tools~\cite{schick2023toolformer, yao2022react} to edit code, run shell
commands, query databases, and browse the web to complete multi-step tasks
autonomously~\cite{claudecode,openaicodex,yang2024sweagent}. Unlike a chat, an agentic
task is a long-running {program} of many LLM calls, each followed by a tool
use whose output is appended to the context. Executions of tens to hundreds of LLM calls
are common. In production traces~\cite{tracelab} that include Claude
Code~\cite{claudecode} and Codex~\cite{openaicodex} sessions, the $90$th-percentile
session issues over $100$ model calls ($241$ for Codex), and its context peaks at $186$K tokens
($235$K for Codex). Even on frontier mixture-of-experts models with compact attention, the
result 
is a large memory footprint---e.g., we compute that such a session holds $36$-$45$\,GB of 
context with
Qwen3-235B-A22B~\cite{qwen3}  and $13$-$17$\,GB   with DeepSeek-V3's latent
attention~\cite{deepseekv3}. 

\begin{figure}[t]
    \centering
    \includegraphics[width=0.9\linewidth]{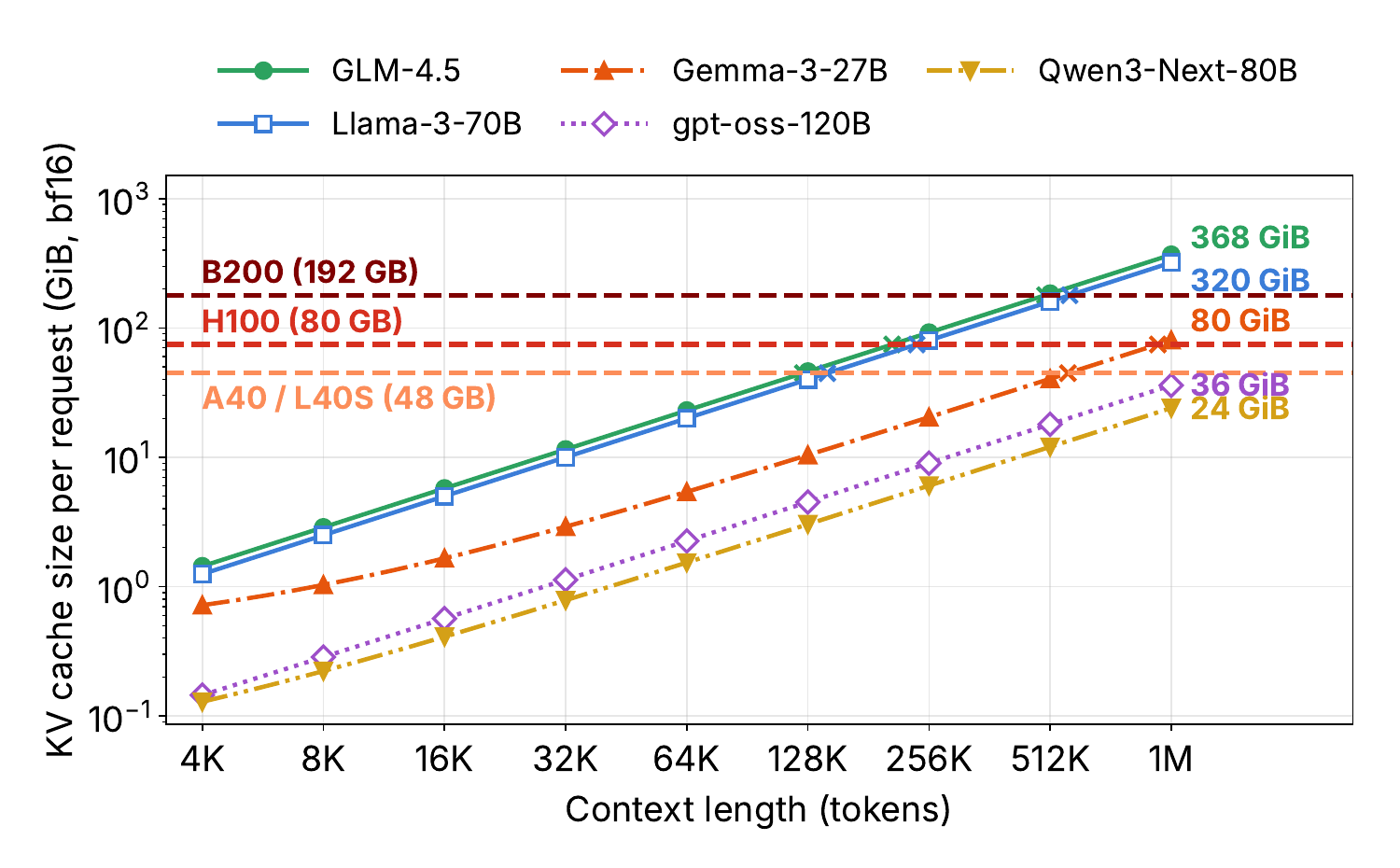}
    \vspace{-5mm}
    \caption{KV cache size of a  request vs.\ context length (assuming bf16 data types) for different models.}
    \label{fig:Context-growth}
    \vspace{-3mm}
  \end{figure}

A request's context is stored in the KV cache~\cite{pope2023efficiently} 
in the accelerator's HBM memory. As the request runs,
it accumulates context, which can consume substantial memory. Figure~\ref{fig:Context-growth}
shows how the increase in the context length of a request in tokens (X-axis) increases the
KV cache size (Y-axis). The figure depicts lines for multiple models, and shows the
KV cache size for 1M tokens (upper right side numbers). The figure also shows, in horizontal
lines, the size of the HBM memory in popular machines. We can see that one  
GLM-4.5~\cite{glm45} request with a $128$K-token context fills 40 GB; two of them would
fill 80GB, which is the memory of an H100 GPU. 


To serve multiple requests simultaneously, servers continuously batch requests based on specific scheduling policies \cite{zheng2024sglang, kwon2023pagedattention, yu2022orca} and availability of space in the KV cache.
A request is batched only if its context fits in the KV cache. Agentic requests hold so much memory that, within a few requests, the KV cache saturates and the batch size freezes, leaving the GPU's compute underused while requests wait in queues. 

To avoid this problem,
KV caches can be periodically summarized to shrink their size.
A summarizer uses the record of the agent's history to read or recompute the KV cache and then 
summarize it.
Recent work shows that frontier agents that compact their history whenever it reaches $12$-$18$K tokens and restart from the summary end up losing little to no task accuracy~\cite{belcak2026longcontext}. That work, however, reports only final accuracy, not whether compressing and restarting changes the agent's trajectory and with it the cost to serve.  

\begin{figure}
    \centering
    \includegraphics[width=1\linewidth,trim={0 0 8mm 0},clip]{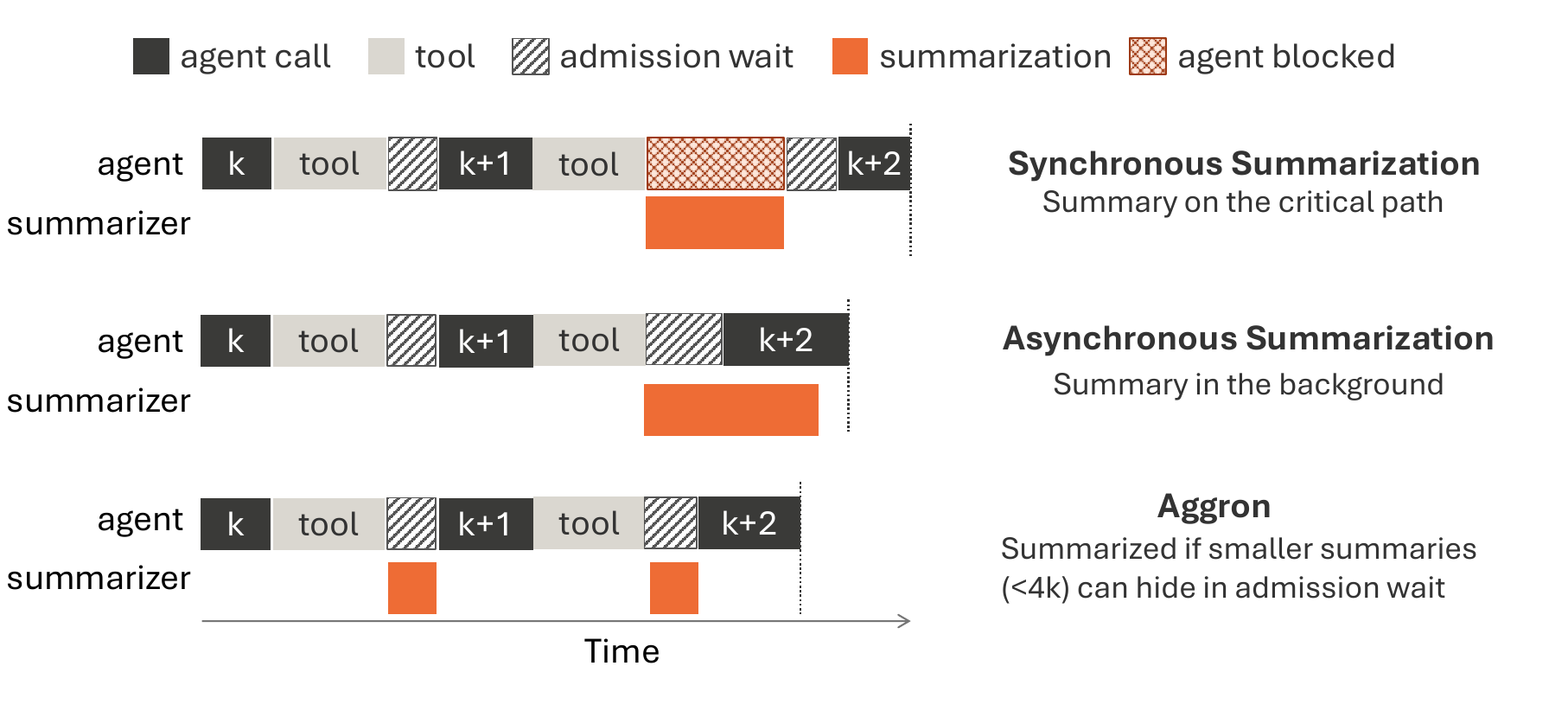}
    \vspace{-8mm}
    \caption{Comparing different summarization strategies.}
    \Description{Design comparisons of summarization strategies}
    \label{fig:Compression_design}
    \vspace{-2mm}
\end{figure}

An open problem is how to simultaneously serve agentic and summarization requests 
under load on a shared server. In general, there are two approaches: 
synchronous and asynchronous. 
In synchronous summarization, the agent harness waits 
until the context approaches a certain threshold, and then stalls the agent while a summarization request executes. 
In asynchronous summarization~\cite{slipstream2026,apec2027}, the harness
lets the agent 
continue while a request summarizes the context. The agent runs on
the full long context until the summarization completes.
With synchronous summarization, the summarization is in the critical path, raising
the mean and P99 latency. With asynchronous summarization, the summarization and agentic requests compete for memory and compute capacity. The top and middle rows of Figure~\ref{fig:Compression_design} 
show a timeline of synchronous and asynchronous summarization, respectively. 
In the latter, the
k+2 agentic execution takes longer due to contention. In both cases, 
summarization can add significant overhead.

To maximize the cost-effectiveness of summarization, this paper
presents a new serving system called {\em Aggron}. Aggron reduces the overhead
of summarization through several techniques. 
First, Aggron summarizes
agentic requests at every turn, to limit the work performed per summarization 
operation,
and to opportunistically hide summarization with the agent's waiting time in the serving queue.
In addition, Aggron selectively summarizes only 
certain parts of the KV cache, namely those that are likely to
have less impact on the accuracy. Second, Aggron
launches summarization requests from a separate {\em Summarization
Queue}. A summarization is launched
while the corresponding agentic request whose KV cache it
wants to compress is stalled in the main request queue due to 
head-of-line stalls. The Summarization queue adds modest overhead
because it utilizes compute and memory
capacity left over by the head-of-line stalls in the main queue. Third, to maximize efficiency, the {\em server} issues a summarization only
when it predicts that it can complete before the corresponding queued
agentic request will finally execute. The server has information about
waiting agentic requests and unused capacity.  When there is no queue to 
hide behind, the
server reverts to synchronous summarization. With Aggron,
the resulting timeline of the summarization is shown in the bottom
row of Figure~\ref{fig:Compression_design}. We can see that
the execution time decreases.

We measure Aggron's accuracy on real agentic tasks from four benchmarks, served on
NVIDIA H100 NVL GPUs, and evaluate its P99 latency and throughput
with a simulator calibrated against
the same hardware. We use simulations to limit the cost of the many runs needed.
Across three models (Qwen3.6-35B-A3B~\cite{qwen36_35b_a3b}, Gemma-4-26B-A4B~\cite{gemmateam2026gemma4}, and GLM-4.7-Flash~\cite{glm45}), Aggron's accuracy stays within run-to-run noise of uncompressed execution on every benchmark (i.e., within $2.7$\,pp).
Compared to the Slipstream~\cite{slipstream2026} and APEC~\cite{apec2027}
state-of-the-art asynchronous summarization schemes, Aggron reduces P99 latency by $27\%$ and $31\%$, reduces mean latency by $11\%$ and $15\%$, and increases throughput by $1.34\times$ and $1.40\times$, respectively.

This paper makes the following contributions:

\noindent $\bullet$
A characterization of agentic serving under KV pressure, showing that requests spend most of their latency waiting for admission, and that selective summarization
     does not hurt accuracy.

\noindent $\bullet$     
Aggron, the first system to compress agentic requests opportunistically at the server in online, multi-tenant serving.

\noindent $\bullet$  
An evaluation of Aggron's accuracy, trajectory divergence, and 
    serving performance against state-of-the-art techniques on different models.

%% file: sections/2.background.tex
LLM inference runs in two phases~\cite{vaswani2017attention}. \emph{Prefill}
processes the whole prompt in one forward pass and is compute-bound;
long prompts are often split into chunks and interleaved with decode
work~\cite{agrawal2023sarathi}. \emph{Decode} then generates one token per step
and is bound by the bandwidth needed to read the cached state from
HBM~\cite{pope2023efficiently,sheng2023flexgen}. Both phases store, for every
token and every attention layer, a key and a value vector, the \emph{KV cache},
which is kept for the request's lifetime so that later steps need not recompute
it.

\begin{figure}[t]
  \centering
  \includegraphics[width=0.6\linewidth]{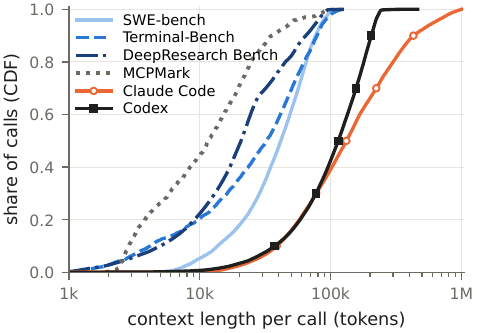}
  \vspace{-3mm}
  \caption{Distribution of context length per agent call, across our benchmarks (Qwen3.6-35B-A3B) and
production traces.}
\Description{Context length per agent call.}
  \label{fig:ctx_cdf}
  \vspace{-3mm}
\end{figure}

Agentic requests make this cache large, since every call carries the task's whole
history. In production traces~\cite{tracelab}, more than half of all
calls carry over $100$K tokens ($8$-$30$\,GB of KV; Figure~\ref{fig:ctx_cdf}).

Architectures have been modified to reduce the per-token cost. Grouped-query
attention~\cite{ainslie2023gqa} shares keys and values across heads, latent
attention~\cite{deepseekv3} compresses them, and linear
attention~\cite{katharopoulos2020linear} replaces them with a fixed-size
recurrent state in some layers. Qwen3.6~\cite{qwen36_35b_a3b} combines GQA with
linear attention in three of every four layers. These reduce the slope of the
footprint but not its growth with context (Figure~\ref{fig:Context-growth}). Even
on Qwen3.6, about $40$ requests of $64$K tokens, a common agentic size, fill an
H100's HBM.

\begin{figure*}[t]
  \centering
  \begin{minipage}[t]{0.49\linewidth}
    \vspace{0pt}
    \centering
    \includegraphics[width=\linewidth]{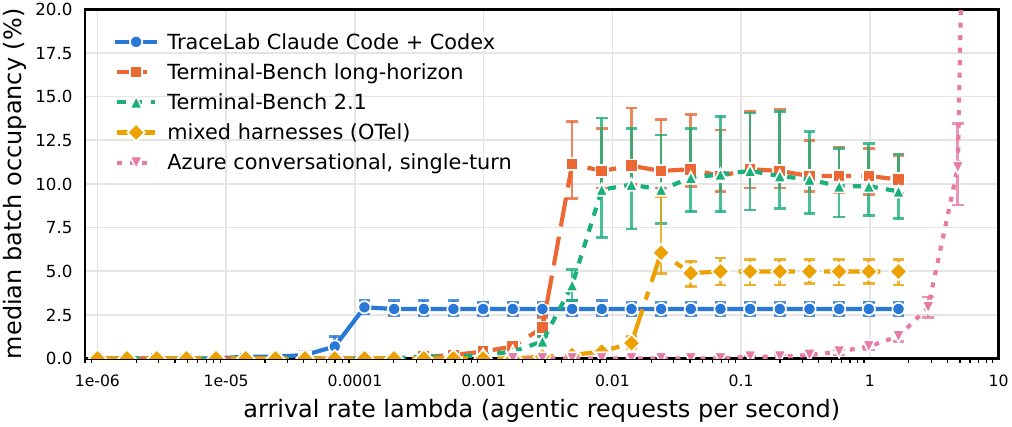}
    \vspace{-7mm}
    \caption{Median batch occupancy versus arrival rate. Single-turn requests
(Azure) are small, and the batch reaches its $1024$ limit. Agentic requests
fill the HBM first.}
    \label{fig:fill_single_turn}
    \Description{Median batch occupancy}
  \end{minipage}\hfill
  \begin{minipage}[t]{0.49\linewidth}
    \vspace{0pt}
    \centering
    \includegraphics[width=\linewidth]{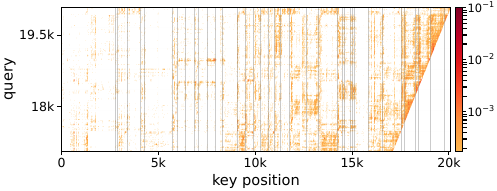}
    \vspace{-6mm}
    \caption{Attention of the last $3$K query tokens of a $38$-iteration SWE-bench
trajectory (astropy-14309, $20$K tokens) over all earlier tokens, at one head of
a full-attention layer (layer $15$) of Qwen3.6-35B-A3B; log color scale}
    \Description{Attention of the last 3K query tokens of a 38-iteration SWE-bench
trajectory (astropy-14309, 20K tokens) over all earlier tokens, at one head of
a full-attention layer (layer 15) of Qwen3.6-35B-A3B; log color scale.}
    \label{fig:attn}
  \end{minipage}
\end{figure*}

Serving systems batch requests at every forward pass, adding new requests and
retiring finished ones (\emph{continuous batching}~\cite{yu2022orca}).
A waiting request is admitted, in first-come-first-served order, only when enough
blocks are free to hold its context; KV of a shared prefix can be reused from a
prefix cache instead of being prefilled again~\cite{zheng2024sglang,lmcache}.
For agentic requests, KV capacity, not compute, decides how many requests run
together. We replay the long traces~\cite{tracelab, tbench21dsh2026,
t1longhorizon2026, exgenticagenttraces2026} and single-turn
traces~\cite{splitwise2024, azurellmtrace2023} and measure batch occupancy
(Figure~\ref{fig:fill_single_turn}) at tensor parallelism of $8$. Single-turn
requests are small, and the batch reaches its limit of $1024$. Agentic requests
fill the HBM first, and the batch plateaus well below it, at a median of $37$
requests on production traces, while the rest wait in the queue until enough KV is free.

A request's footprint can also be cut at runtime by compressing its context.
\emph{Pruning}~\cite{zhang2023h2o, liu2023scissorhands, yang2024pyramidinfer,
chen2024nacl, xiao2024streamingllm} drops tokens, or their cached keys and
values, chosen by position or by the attention they have received; it edits the
cache in place and needs no recomputation, but evicting a token in the middle of
a sequence breaks prefix-cache matches for every block after it.
\emph{Summarization}~\cite{tcrallm2023,acon2025,retrac2026,slipstream2026}
replaces spans of the context with text generated by the model itself; it costs
an LLM call, and the KV of the rewritten context must be prefilled again.
Deployed agents summarize \emph{synchronously}, stopping when the context nears
the model's window.

%% file: sections/3.PA.tex
To understand the impacts and challenges of summarization on agentic requests, we start by running experiments on an H100 GPU with $94$\,GB of HBM3, using the Qwen3.6-35B-A3B-FP8 \cite{qwen36_35b_a3b} model, served by a vLLM \cite{kwon2023pagedattention} server. We draw agentic requests from the Terminal-Bench \cite{merrill2026terminal}, MCPMark \cite{wu2025mcpmark}, DeepResearch Bench \cite{du2025deepresearch}, and SWE-bench \cite{swebench2023} benchmark suites. We present our main observations that motivate Aggron's design by answering the following questions.

\paragraph{\bfseries\itshape What is the best way to reduce the memory footprint of a request?}
Pruning is the cheaper of the two runtime methods, but it is the wrong fit for
agents, because it decides what to keep from the attention a token has received
so far, and agents return to old context. Figure~\ref{fig:attn} shows the
attention of the last $3$K query tokens of a $38$-iteration SWE-bench trajectory.
Vertical stripes across the whole history are spans, read thousands of tokens
and many iterations earlier, that later steps keep attending to. Worse, the history is revisited unpredictably, with keys around
$6$-$8$K receiving almost no attention from most later queries, yet two later
iterations return to them. An attention-ranked budget would already have evicted exactly those tokens.
Offloading them to host memory keeps them reachable, but every decode step then
fetches KV over the CPU--GPU interconnect, which has
$10$-$50\times$ less bandwidth than HBM~\cite{lee2024infinigen,sun2025shadowkv}.

Summarization decides what to keep from meaning instead, with the model itself
as the judge, and it is safe at small budgets. Agents whose prose is capped at
$4$K tokens lose at most $2.7$\,pp of uncompressed accuracy,
consistent with frontier agents compacted to $16$K~\cite{belcak2026longcontext}.
We therefore compress
by summarization.

\takeaway{
        {Pruning discards context that agents later revisit; summarization keeps what
  matters.}
  }

\begin{table}[t]
  \centering
  \caption{Accuracy of whole-context summarization at different trigger lengths
(Qwen3.6-35B-A3B, $50$ DRB and $100$ SWE-bench tasks).}
\vspace{-3mm}
  \label{tab:ctx-accuracy}
  \begin{tabular}{lccccc}
    \toprule
    \textbf{Benchmark} & \textbf{4K} & \textbf{16K} & \textbf{32K} & \textbf{64K} & \textbf{120K} \\
    \midrule
    DRB & 41.2\% & 42.3\% & 38.8\% & 42.2\% & 42.2\% \\
    SWE & 58.0\% & 60.0\% & 65.0\% & 63.0\% & 64.0\% \\
    \bottomrule
  \end{tabular}
  \vspace{-2mm}
\end{table}

\paragraph{\bfseries\itshape Are all parts of a request's memory compressible?}

Summarizations can compress the entire context or parts of it. To ensure that requests can run indefinitely, existing systems trigger summarizations whenever the context approaches the maximum model length. We experiment with summarization triggers and observe their impact on accuracy. The trigger refers to the context size in token count after which compression through summarization occurs, effectively fixing the maximum context size. Table \ref{tab:ctx-accuracy} tabulates the variation in accuracy with the trigger size. SWE-bench consists of coding tasks whose context includes code snippets, among other information. Summarizing raw code results in loss of variable names and assertion statements, as the summary captures the function's intent but not its semantics. This results in incorrect decisions later in the trajectory, hurting accuracy. The accuracy saturates after $32$K as most requests do not exceed a $32$K context size.

On the other hand, the context of requests from DeepResearch Bench (DRB) is populated with results from web searches. These are verbose, and when given a context, summaries effectively keep only relevant information. We thus see a relatively stable accuracy trend on DRB while accuracy drops by up to $6$\,pp on SWE, despite having an uncompressed context size of $40$-$50$K tokens. The distinction in data type, produced by tools, helps us identify suitable summarization targets. We use the term \emph{prose} to refer to data that can be compressed without hurting accuracy, and \emph{non-prose} for incompatible summarization targets. Aggron opts to selectively compress prose data when it gets large, while ensuring non-prose is kept verbatim.

\takeaway{
     {Prose data are suitable targets to summarize, while non-prose data are not suitable for summarization.}
  }

\begin{figure*}[!t]
  \centering
  \begin{subfigure}[b]{0.3\linewidth}
    \centering
    \includegraphics[width=\linewidth]{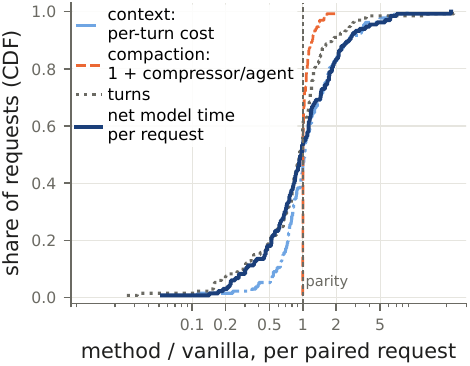}
    \caption{}
    \label{fig:cdf}
  \end{subfigure}\hfill
  \begin{subfigure}[b]{0.34\linewidth}
    \centering
    \includegraphics[width=\linewidth]{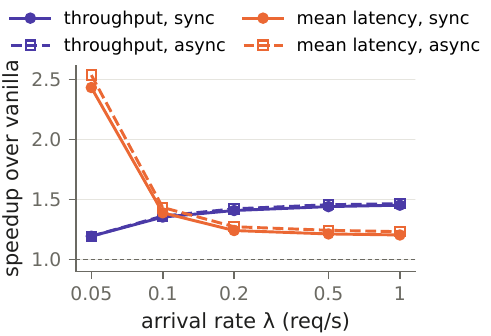}
    \caption{}
    \label{fig:mot_tpt_p99}
  \end{subfigure}\hfill
  \begin{subfigure}[b]{0.36\linewidth}
    \centering
    \includegraphics[width=\linewidth]{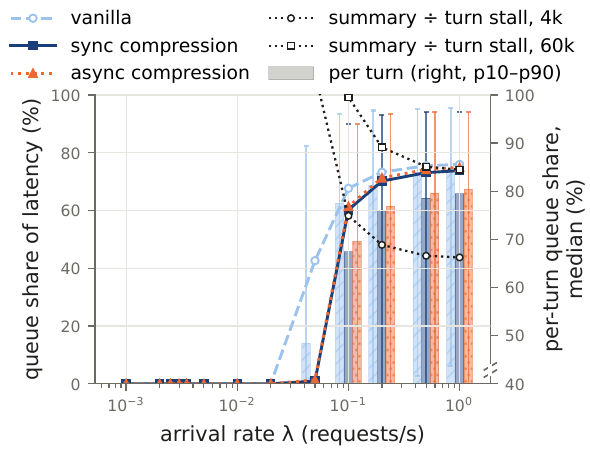}
    \caption{}
    \label{fig:Queue share}
  \end{subfigure}
  \vspace{-7mm}
  \caption{Synchronous and asynchronous summarization at the $4$K prose cap on our benchmark traces (SWE-bench, Terminal-Bench, MCPMark).
  (a) Distribution of per-request latency normalized to vanilla.
  (b) Throughput and mean program latency over vanilla.
  (c) Share of latency spent queuing.}
  \Description{Three panels: a CDF of per-request latency normalized to
  vanilla; throughput and mean program latency over vanilla for synchronous and
  asynchronous summarization; and the share of latency spent queuing for vanilla,
  synchronous and asynchronous summarization.}
  \vspace{-3mm}
\end{figure*}

\paragraph{\bfseries\itshape What does compressing only prose buy?}

Compressing only prose shrinks each request less than compressing everything,
yet the system still gains. On the coding-heavy production traces, where prose
is a small share, the $4$K cap lowers KV memory use by $1.35$-$1.57\times$
(Table~\ref{tab:production}), and how much that helps depends on load. At low
load, nothing queues, and shorter contexts make each step only slightly cheaper
($1.02\times$ mean latency). Near the knee ($0.05$-$0.1$~req/s), the
smaller footprint keeps requests from queuing, and mean latency improves most
($1.31$-$1.33\times$). At the highest rates, both arms queue; the smaller footprint
admits more requests per batch, holding throughput $1.07\times$ higher, and P99
improves $1.17\times$.

With synchronous summarization, if a request's prose exceeds
the cap, the request pauses, the same model summarizes the prose, and the
request resumes. The summarization call sits on the
critical path, time the request would otherwise spend on its task. In exchange,
every later turn runs on a shorter context. The new context can also change the
trajectory, so the agent may take more or fewer turns to finish. Figure~\ref{fig:cdf} shows the
distribution of the request latency normalized to its corresponding vanilla run. We see that the median effect is near parity. The CDF weights every program equally, so the many short programs, which rarely compact, hold the median at parity, while the long programs that dominate GPU time gain most. Figure~\ref{fig:mot_tpt_p99} shows that
from $0.05$~req/s up, synchronous summarization raises throughput
$1.19$-$1.45\times$ over vanilla, from $1.05$-$1.5\times$ larger mean batches.
Asynchronous summarization matches it in throughput and gains little in latency. This runs counter to prior
findings that asynchronous summarization outperforms
synchronous~\cite{slipstream2026,apec2027}, which hold when summarization is a large
share of latency, and the background summary does not compete with other requests
for the GPU. The reduced summarization latency is only a fraction of the stalled time it sees every turn, as seen in
Figure~\ref{fig:Queue share}. It shows that once memory saturates, requests spend over
$60\%$ of their end-to-end time queued under every method, due to head-of-line stalls. With end-to-end latency determined by queue time, the summary's own runtime is a small share of latency under either scheme, so hiding it buys little.

\begin{table}[t]
\centering
\small
\caption{Gain of the $4$K prose cap over vanilla on production traces~\cite{tracelab}.}
\vspace{-3mm}
\begin{tabular}{lcccc}
\toprule
$\lambda$ (req/s) & Throughput & Mean latency & P99 latency & HBM use \\
\midrule
0.01 & 1.00$\times$ & 1.02$\times$ & 1.03$\times$ & 1.57$\times$ \\
0.05 & 1.06$\times$ & 1.33$\times$ & 1.07$\times$ & 1.50$\times$ \\
0.1  & 1.07$\times$ & 1.31$\times$ & 1.08$\times$ & 1.35$\times$ \\
0.5  & 1.07$\times$ & 1.19$\times$ & 1.16$\times$ & 1.37$\times$ \\
1    & 1.07$\times$ & 1.17$\times$ & 1.17$\times$ & 1.37$\times$ \\
\bottomrule
\end{tabular}
\label{tab:production}
\vspace{-3mm}
\end{table}

\takeaway{
  Under load, a summary's cost is the queue wait it adds, not its own runtime.
}

Table~\ref{tab:hol-4k} reports queue statistics under the $4$K prose cap. At high loads, the request at the head of the queue cannot be admitted for $75$-$89\%$ of the
run. In about $5\%$ of that blocked time, the head would have fit had it been slightly smaller. Compressing its prose tail, which frees $\sim 2.4$K tokens, would have admitted it. Although small
as a share, this time is costly, because the entire queue stalls behind a single request. At
$\lambda = 0.2$, it amounts to $15$ minutes of a $6$-hour run with a mean queue of $\sim 250$ waiting
requests. Admission also does not stop at the head. Once it is admitted, the next request becomes
the head and is checked in turn. Had every waiting request's tail been compressed, at least one
more request would have been admitted in $\sim 46\%$ of the blocked time. This motivates
compressing waiting requests before they reach the head of the queue.

\begin{table}[t]
\centering
\small
\caption{Queue statistics under the $4$K prose cap. }
\vspace{-3mm}
\begin{tabular}{lccc}
\toprule
$\lambda$ (req/s) & Head blocked & Compressed tail admits  & $\geq$1  admitted \\
  & (\% of run) & (\% of blocked) & (\% of blocked)  \\
\midrule
0.05 & 8.7\%  & 4.8\% & 12.8\% \\
0.1  & 74.8\% & 4.5\% & 44.0\% \\
0.2  & 83.0\% & 4.9\% & 46.0\% \\
0.5  & 88.7\% & 5.0\% & 46.5\% \\
1    & 86.2\% & 5.0\% & 46.2\% \\
\bottomrule
\end{tabular}
\label{tab:hol-4k}
\vspace{-2mm}
\end{table}

\takeaway{
     {Opportunistically compressing the prose tail end can admit the blocking head and another request.}
  }

%% file: sections/4.aggron.tex
Based on the previous observations, we propose Aggron, an agentic serving
system that compresses agentic contexts aggressively and hides the resulting
summarization latency inside queuing delay. Aggron is based on four main ideas.

\noindent\textbf{1. Aggron compresses only the prose regions of a request's context.}
Tool output carries structure as content, the identifiers the agent acts on next,
while natural-language reasoning compresses freely (Observation 1, Observation 2). Aggron classifies
every observation and makes only the prose a candidate for the compressor.

\noindent\textbf{2. Aggron hides summarization latency in admission wait.}
Under load, a summary's cost is the time its request spends waiting (Observation 3). A
request returning from a tool call is not running but waiting for readmission, so
Aggron issues the summarization into that wait, and the summary is ready when the
request is admitted.

\noindent\textbf{3. Aggron serves summarization requests from a separate queue.}
Small thresholds produce many summaries, and in the main queue they take batch
slots from agentic requests (Observation 3). Aggron's scheduler
draws from the summarization queue only when no further agentic request can be
admitted.

\noindent\textbf{4. Aggron treats summarization as a server-issued mechanism.}
Compressing the prose tails of waiting requests can admit the blocked head and
the request behind it (Observation 4), so Aggron decides when to compress from the queue
state, not only from a request's context length.

\subsection{Design overview}
\label{sec:design:overview}

Figure~\ref{fig:aggron} provides an overview of the Aggron architecture and the path a
request takes through it. A request enters the Session Registry
\circled{1}, which holds the per-session state, namely the frozen summaries, the
protected region, and the compressible accumulation $K$. Requests returning
from a tool call re-enter through the same point \circled{7}, where the
Context Classifier partitions the new observations into protected and
compressible. Prose increments $K$ by its token count, calculated from its transcript. A non-prose
observation advances the protected region over everything before it, including
any prose not yet compressed, and resets $K$ to zero; $K$ also resets when a summarization fires. The agentic request is forwarded to the queue \circled{2}. The
session is also handed to the Controller \circled{3}, which reads the
request's position in the agentic queue \circled{4} and the admission rate \circled{6} from the admission control, and decides if a request's tail should be compressed.
When $K$ exceeds a threshold $\theta$ that falls the deeper the request sits in the queue
(Section~\ref{sec:design:trigger}), and three gates confirm that the summary lands before the
request reaches the head, shrinks the tail enough to be worth it, and frees at least one KV
block (Section~\ref{sec:design:masking}), the Controller enqueues a summarization for that
session on the summarization queue \circled{5}. Once the summarization completes, the Registry
replaces the compressed region of that request's
context in place, in the agentic queue; the request never left the
queue and keeps its position. If the summarization has not landed when the
request reaches the head, the request holds the head until it does.
Admission Control draws from the agentic queue in default order, and from
the summarization queue only when the agentic head does not fit in free KV space,
and dispatches to the GPU \circled{8}. A preempted request returns to the head
of the agentic queue with its KV discarded \circled{9}.
\input{figures/Aggron/fig_aggron}
\subsection{Identifying compressible context}
\label{sec:design:classify}

\textbf{Classification.} Aggron classifies at the granularity of an
observation, i.e.\ a single tool result in the transcript. For each
observation, Aggron recovers the call that produced it by its
\texttt{tool\_call\_id} and inspects the tool name and arguments. When the
agent reads the same resource again, such as the same file, the new
observation replaces the earlier one, so the context holds one copy.
Aggron classifies as non-prose the output of three families of tools,
(i) file-system and shell inspection (\texttt{cat}, \texttt{grep},
\texttt{read\_file} and similar), (ii) database access (\texttt{execute\_sql},
schema listings, query plans), and (iii) program and test execution
(\texttt{pytest}, \texttt{python}). A SQL result set or a test log, with its
pass/fail counts and tracebacks, is structured content in the same way a file
read is. The model's reasoning, the tool calls themselves, and web search and
fetch results are prose and remain compressible.

\textbf{The incompressible region.} A non-prose observation protects not only
itself but everything before it that has not yet been compressed. When one
arrives, the protected region advances to cover the transcript up to and
including it, so it stays a contiguous prefix and never separates a call from
the result it explains. The price is that prose written before a file read or
a test run is never summarized. Compressible prose is therefore exactly the
prose that follows the most recent non-prose observation.

\subsection{Triggering summarization}
\label{sec:design:trigger}

\textbf{Counting just the prose.} Aggron maintains a per-request counter $K$
of the tokens that summarization can actually reduce. Writing $\mathcal{M}$ for
the messages in the context, $t(\cdot)$ for a token count, and $\mathcal{H}$,
$\mathcal{F}$, $\mathcal{P}$ for the head, the frozen
summaries, and the protected region, respectively,
\begin{equation}
K \;=\; \sum_{m \in \mathcal{M}} t(m)
   \;-\!\!\!\sum_{\;m \,\in\, \mathcal{H} \cup \mathcal{F} \cup \mathcal{P}}\!\!\! t(m).
\label{eq:accum}
\end{equation}
$K$ is therefore not the context length. It excludes exactly the regions
summarization cannot shrink, and counts only the prose tail, i.e.\ the prose that
follows the most recent non-prose observation. A non-prose observation moves
everything before it into $\mathcal{P}$ and so returns $K$ to zero; so does a
summarization.

\textbf{Limiting the prose tail.} Aggron keeps a hard backstop. When $K$
exceeds the default budget $T_{\mathrm{def}}$, summarization fires on the agent's
critical path, as in any summarization scheme. This upper limit guarantees that
summarization does occur and that the compressible tail stays bounded. Firing only
there, however, wastes the one resource the workload has in abundance, the queuing time. A
request that has returned from a tool call sits in the waiting pool for a
readmission delay that it pays whether or not anything runs on its behalf
(Section~\ref{sec:design:masking}), and a summarization issued into that window is
hidden. Aggron therefore also compacts {preemptively} while a request
waits, at a second and much lower threshold.

\textbf{Queue depth-dependent thresholding.} A request at the front of the queue will be
admitted shortly and should be left alone; one at the back has time to spare
and can be compacted hard. Aggron makes the threshold a function of
position. Let $a$ be the request's age rank among live programs ($a = 1$
oldest), $f$ its depth in the queue ($f = 1$ at the back), and $s$ its
context-size rank ($s = 1$ largest), each normalized to $[0,1]$. With a weight
$w$ on the size term,
\begin{equation}
g \;=\; \frac{a\,f\,\bigl(1 - w + w\,s\bigr)}{1+n},
\qquad
\theta \;=\; T_{\mathrm{hi}} - g\,\bigl(T_{\mathrm{hi}} - T_{\min}\bigr),
\label{eq:theta}
\end{equation}
where $T_{\mathrm{hi}} = T_{\mathrm{def}} - \epsilon$ and $n$ counts the program's queue-time
summaries since its last $4$K-cap compaction, so each further one pulls $\theta$ back toward
$T_{\mathrm{hi}}$. Aggron compacts a waiting request when $K > \theta$. The threshold runs from
just under $T_{\mathrm{def}}$ at the head of the queue down to $T_{\min}$ at the back, so
$\theta$ is an aggressiveness dial that the queue itself sets. Deep in the queue, Aggron
compresses a tail far smaller than any critical-path scheme would consider, and at the head
it does nothing. We set $T_{\min} = 1024$,
$T_{\mathrm{def}} = 4096$, $\epsilon = 64$ tokens, and $w = 0.5$, which weighs
waiting and context size equally. Among requests equally deep in the queue,
Aggron compacts the larger contexts first, since they return the most KV to
the pool. The decision is taken afresh on every queue visit and at most once
per visit, since both $K$ and the request's position change while it waits.

\textbf{Avoiding re-summarization.} When the trigger fires, Aggron
compresses the prose tail, {appends} the result to the context as a frozen
segment, and resets $K$ to zero. It does not feed the standing summary back
into the next compressor's input. The distinction is a main reason Aggron
can compress aggressively without losing accuracy. Under the conventional
scheme, the summary is re-summarized at every summarization, so its information is
distilled further every time. Freezing passes every token through the
compressor exactly once and pays in linear context growth instead, still well
below the growth of an uncompressed context. Freezing has a second,
systems-side benefit by keeping the KV prefix
valid for the whole episode. The conventional scheme rewrites the summary at
every summarization and forces a full re-prefill of the context.

\textbf{Upper-bounding the context.} Linear growth reaches the model window eventually.
Aggron therefore maintains a reset ceiling $R$, as agent harnesses do~\cite{claudecode,openaicodex}. When the assembled context
exceeds $R$, Aggron summarizes everything, including the frozen segments, into a
single segment and restarts. Aggron sets $R$ just below the max-model length
to give a margin for the completion budget. The protected set is folded in at
a reset, so the reset is the one point at which Aggron does compress non-prose;
it is reached rarely, and the alternative is a request that cannot be served at
all.

\subsection{Masking summarization latency}
\label{sec:design:masking}

Aggron's observation is that under load an agentic request spends most of its lifetime not running; in fact, at high loads, $60$-$79\%$ of end-to-end latency is spent
waiting for admission (Figure~\ref{fig:Queue share}). Aggron spends it on summarization.

\textbf{Issuing a summarization request.} When a request returns
from a tool call and its compressible accumulation $K$ exceeds $\theta$, and passes a gate,
Aggron enqueues the summarization immediately, rather than waiting for the agent
to reach the next LLM call and discover that it must compact first. The
summarization and the queuing delay then overlap. If the summary completes
before the request reaches the head of the queue, the stall is fully hidden,
and the agent observes none of it. If it does not, the request keeps its
position and holds the head until the summary lands. Aggron does not admit it
uncompressed, which would discard the summarization's benefit. Holding the head delays this one request, but the
requests behind it wait too, and they spend that time completing their own
summarizations. The hideability gate below keeps this case rare.

\textbf{Gating the summarization queue.} Crossing $\theta$ makes a request a candidate, not a
summarization. Aggron issues it only when three gates pass.

\textbf{Gate 1. The summarization is hideable.} Aggron predicts the summarization's latency from the
summarizations it has just served. Using the $90$th percentile of the last $50$ measured
issue-to-landing times per summary token, which include their wait in the summarization queue,
it scales by this summarization's summary cap $S_{\max}$. Normalizing by length matters
because summaries vary widely in size, and a single percentile under-predicts the long ones.
A summarization that lands late raises the estimate, so later decisions learn from this. To
handle cold starts, the first ten summarizations fall back to a model of the summarization's own
cost, $t_{\text{sum}} = a_{\text{prefill}} K + S_{\max}\,t_{\text{step}}$ (one prefill of the tail
plus the decode of an $S_{\max}$-token summary), plus the backlog of the summarization queue ahead of it,
drained at the recent summary admission rate. It estimates the request's own wait as its
position in the queue divided by the recent admission rate, and issues the summarization
only if that wait covers the prediction.

\textbf{Gate 2. The summarization is worth it.} The summary is appended to the prompt
permanently, so compressing a tail barely larger than its summary gains little. The
summary's size is not known until it is decoded, so Aggron bounds it by the summarizer's
output cap $S_{\max}$ and issues a summary only if $K \geq \alpha\,S_{\max}$. The summary then
holds at most $1/\alpha$ of the tail it replaces. We set $\alpha = 2.5$.

\textbf{Gate 3. The summarization frees memory.} The rewrite must release at least one KV block,
since a summarization that shortens the context without returning a block does nothing for
admission. Aggron requires $\lceil C/b \rceil - \lceil (C - K + S_{\max})/b \rceil \geq 1$,
where $C$ is the request's context and $b$ the block size. Only requests past their first
call are candidates, since a first call has no prose tail yet. A request whose summarization is
still in flight is skipped, and each request is decided once per visit to the queue.

\textbf{Summarization requests occupy their own queue.} Summarization requests
and agentic requests have different shapes, especially since we compress only the small prose tail. A summarization is a single
prefill with a short, capped decode and a small context, while an agentic turn
is a moderate prefill with an open-ended decode over a larger context.
Figure~\ref{fig:fill_single_turn} shows that agentic requests saturate memory
before compute, while single-turn requests, which summarization requests are,
saturate compute with little memory pressure. However, a single queue lets summarization calls take KV and batch slots from agentic requests, which lengthens queuing, as seen in Figure~\ref{fig:Queue share}.  Aggron therefore maintains a
separate summarization queue. The agentic queue is the default queue. When the request at its head does not fit in free KV space, no
agentic request is admitted. This is checked at every batch. That is exactly when the scheduler draws from the
summarization queue, admitting summarization work into the KV and batch slots the
blocked head cannot use, instead of leaving them idle. This is the point in
the cycle at which summarization is most valuable, since the pool is full and a
summarization is the only admission that can fit.

%% file: figures/Aggron/fig_aggron.tex
\begin{figure}[!t]
    \centering
    \includegraphics[width=1\linewidth]{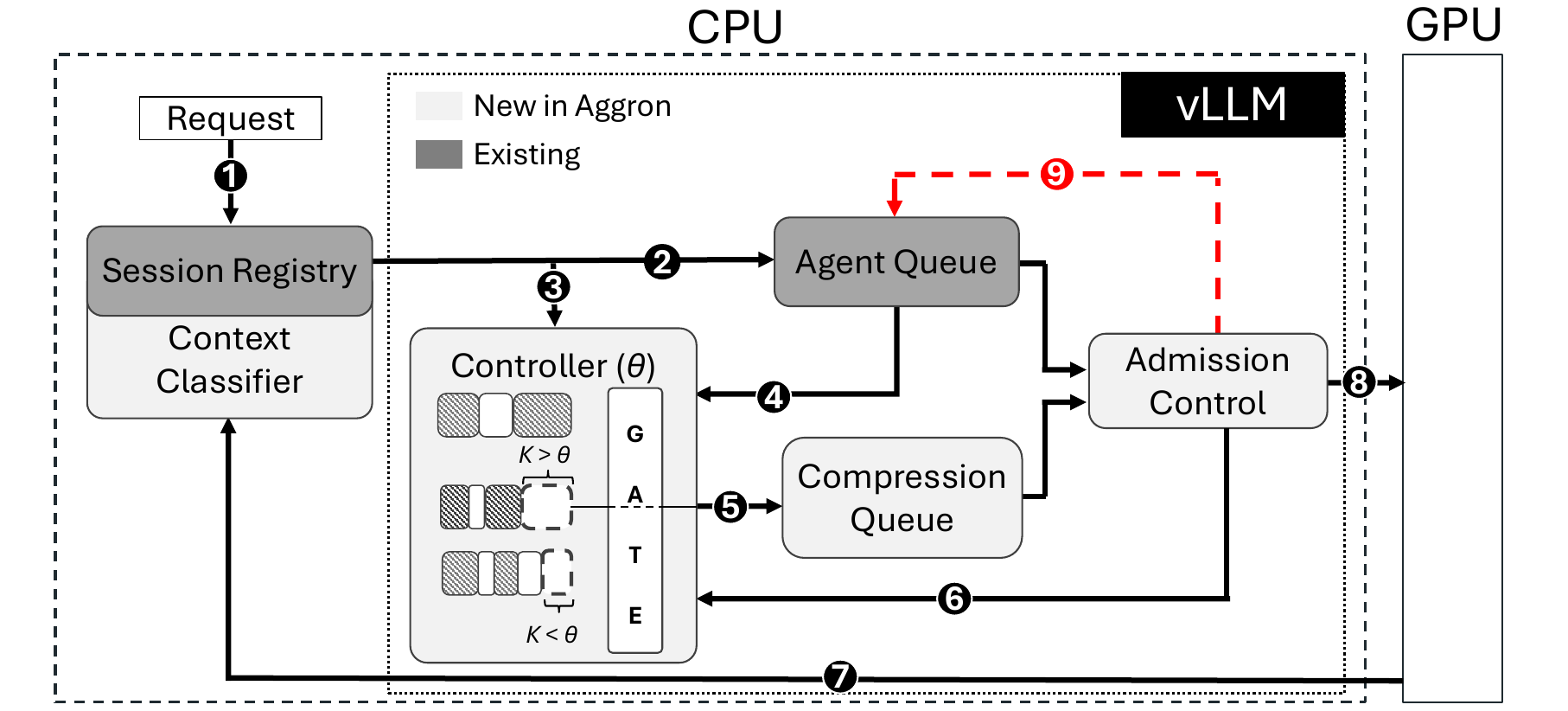}
    \vspace{-7mm}
    \caption{Aggron architecture}
    \label{fig:aggron}
    \Description{Aggron architecture}
    \vspace{-4mm}
\end{figure}

%% file: sections/5.setup.tex
\noindent{\bf Hardware Infrastructure.}
We perform our accuracy analysis on a single NVIDIA H100 NVL GPU with $94$\,GB of
HBM3. The model is served through vLLM with a $131{,}072$-token context window, up
to $256$ concurrent sequences, and $92\%$ of HBM given to the KV cache, yielding a
resident KV pool of roughly $2.49$M tokens in $1056$-token blocks.

\noindent{\bf Models.} We evaluate three models to separate the method from any
single architecture. Qwen3.6-35B-A3B (FP8) \cite{qwen36_35b_a3b} is the primary model and carries the
headline result; GLM-4.7-Flash \cite{glm45} and Gemma-4-26B-A4B \cite{gemmateam2026gemma4} verify that the effect
generalizes. The three span hybrid linear, latent, and grouped-query attention. We report headline serving results on Qwen, the model whose simulator
we validate against live serving (Table~\ref{tab:simval}).

\noindent{\bf Arms.} Aggron compresses a prose tail once it holds between
$T_{\min} = 1024$ and $T_{\mathrm{hi}} = 4096 - \epsilon$ tokens
(Equation~\ref{eq:theta}), at a point that depends on load. The {grow-4k} and
{grow-1k} arms fix that point at its two ends. The serving model summarizes the
prose, and the summary is appended to the prompt and never rewritten, so the cached
prefix survives. Non-prose tool output stays verbatim. An Aggron summary is the same
summary issued while the request waits, so only its timing and threshold differ.
The summarizer writes about $300$-$500$ tokens at any threshold, keeping $10$-$20\%$
of the tail it replaces under grow-1k and $5$-$10\%$ under grow-4k, as one tool
result often carries the tail well past its threshold. Accuracy cannot fall as a summary keeps
more, so the two arms bound Aggron's. {Vanilla} runs without compaction in the full
context window. Slipstream and APEC compact the whole context at $60$K, past the
point where accuracy saturates (Table~\ref{tab:ctx-accuracy}).

\noindent{\bf Noise.} Agentic benchmark scores move by several points across
identical runs~\cite{atil2024nondeterminism,yuan2025numerical,song2025greedy,reliabilitybench2026}.
We therefore run vanilla twice, report the mean, and take the disagreement
between the two runs as the noise band, $1.96\sqrt{D}/n$, where $D$ is the
number of tasks on which the runs disagree and $n$ the number of tasks scored.
Each row is scored on the tasks that every run of that model completed.

\noindent{\bf Trajectory divergence.} Summarization can change how an agent solves
a task, not just whether it succeeds. Following prior
work~\cite{costbench2025,llmconsistency2026,agentevalsurvey2025}, we compare
the ordered sequences of tool calls two runs make on the same task, using Levenshtein distance $d$ normalized by the longer sequence, from $0$ (identical)
to $1$ (nothing shared). We complement it with
Jaccard overlap of the sets of tools used, which ignores order. Because a rerun
already strays, we report the median excess over two vanilla runs,
$\Delta d = d(\text{grow}, \text{vanilla}) - d(\text{vanilla}, \text{vanilla}')$.
$\Delta d = 0$ means summarization changes the trajectory no more than a rerun
does.

\noindent{\bf Applications.}
We evaluate on four agentic benchmarks, each long enough for summarization to engage,
covering code editing, raw shell, web research, and structured tool protocols. \emph{SWE-bench Verified}~\cite{swebench2023} ($100$
stratified instances) covers repository-level bug fixing. The agent edits and tests
inside a per-instance container and is scored by the official harness on whether
its patch resolves the issue. \emph{Terminal-Bench}~\cite{merrill2026terminal}
($49$ tasks) covers open-ended terminal work driven by the benchmark's own terminus
agent, one container per task, scored by task-specific tests.
\emph{DeepResearch Bench}~\cite{du2025deepresearch} ($50$ English queries) covers
long-horizon web research against a self-hosted SearXNG instance, identical
across arms, with reports graded by the RACE rubric judge.
\emph{MCPMark}~\cite{wu2025mcpmark} ($66$ tasks) covers tool use over the Model Context
Protocol across filesystem and PostgreSQL services, run under the benchmark's
own orchestration agent and scored by its per-task verifiers.

\begin{table}[t]
\centering
\small
\caption{Simulated vs.\ live vanilla serving on one H100 (Qwen3.6-35B-A3B; $200$ programs at $\lambda{=}0.05$, $500$ at $0.2$). Error (\%) is (sim $-$ live)/live against the mean of two live runs at the same $\lambda$.}
\vspace{-2mm}
\label{tab:simval}
\begin{tabular}{lcccc}
\toprule
$\lambda$ & Throughput & P50 & P99 & Mean \\
\midrule
$0.05$ & $+4.3$ & $-16.3$ & $-3.0$ & $-14.5$ \\
$0.2$  & $+1.9$ & $+2.1$  & $-3.4$ & $-0.6$ \\
\bottomrule
\end{tabular}
\vspace{-2mm}
\end{table}

\noindent{\bf Simulation Infrastructure.}
The accuracy harness runs agentic tasks sequentially, each a closed loop of model
calls and tool executions, so it cannot exhibit the contention under concurrent
load that our scheduler resolves. We therefore
evaluate the system in a trace-driven discrete-event simulator that replays
recorded agentic trajectories under a synthetic arrival process. Replaying
recorded trajectories is valid because each arm replays its own trajectories
from the accuracy runs, and summarization leaves them within run-to-run variation
(Section~\ref{sec:acc}).
\input{figures/Eval/sota_comp}
We follow Vidur~\cite{vidur}, calibrating a few constants on one GPU and simulating
serving at scale. We build our own simulator because the unit we serve is an agentic
program, not an independent request. A program issues its next call only after the
previous one finishes and its tool returns, so a slow server delays its later arrivals,
and its KV persists across calls, so each call's prefix hit depends on what was evicted
while its tool ran. The simulator replays recorded
trajectories (per-call context length, generated tokens, and tool time) through
a vLLM-style continuous-batching engine over a paged KV-cache block pool, with
prefix caching and CPU KV offload~\cite{lmcache}, following the same design as recent agent-serving
simulators~\cite{agentservesim,vidur-agent,frontier}. Each serving run replays $1{,}000$ programs drawn
from all four benchmarks, arriving as a Poisson process at rate $\lambda$ programs per
second with a fixed seed, so every arm sees the same arrivals. Live runs load Aggron
into unmodified vLLM~$0.24$ as a subclass of its scheduler, and the simulator runs the
same controller code. Section~\ref{sec:scaling} runs the same simulator on four frontier models, T1 (a fine-tuned
Qwen3.5-122B~\cite{qwen3.5}, TP=$8$), DeepSeek-V4-Flash~\cite{deepseekai2026deepseekv4} (TP=$8$),
GLM-5.3~\cite{glm5} (TP=$16$) and Kimi K3~\cite{kimik3} (TP=$32$), all on H100s. Every model replays
the same mix of long-horizon traces, from T1 on LH-TB~\cite{li2026longhorizonterminalbench},
DSV4 on Terminal-Bench~2.1~\cite{merrill2026terminal}, and GLM-5.3 and K3 on Terminal-Bench
Science~\cite{tbscience}. We fit each model's step costs to the measured H100 vLLM tables of
NVIDIA AIConfigurator~\cite{aiconfigurator} ($3.5$-$9.0\%$ MAPE), and the baselines cap each program's context at $120$K tokens, near the $131$K window of our vanilla arm, and at $24$K, within the budgets of Belcak et al.~\cite{belcak2026longcontext}, summarizing in-band and restarting at the cap.

Against measurements, the simulator predicts decode step time within $7.9\%$
and prefill latency within $9.4\%$ mean absolute error. We validate the simulator end-to-end on vanilla, replaying the same programs live via vLLM on a single H100 and comparing each run with its simulated twin (Table~\ref{tab:simval}). Every metric lies within $3.4\%$ at $\lambda = 0.2$. At $\lambda = 0.05$, the box never queues, so latency tracks host speed. Throughput and P99 lie within $4.3\%$, while the median and mean miss by up to $16\%$, as host speed varied by $21\%$ between the two live runs. Aggron extends this validated engine with its controller, running the same controller code as the live scheduler, and Section~\ref{sec:tpt_and_occ} reports its simulated values.

%% file: figures/Eval/sota_comp.tex
\begin{figure*}[!t]
  \centering
  \begin{subfigure}[b]{0.5\linewidth}
        \centering
        \includegraphics[width=1\linewidth]{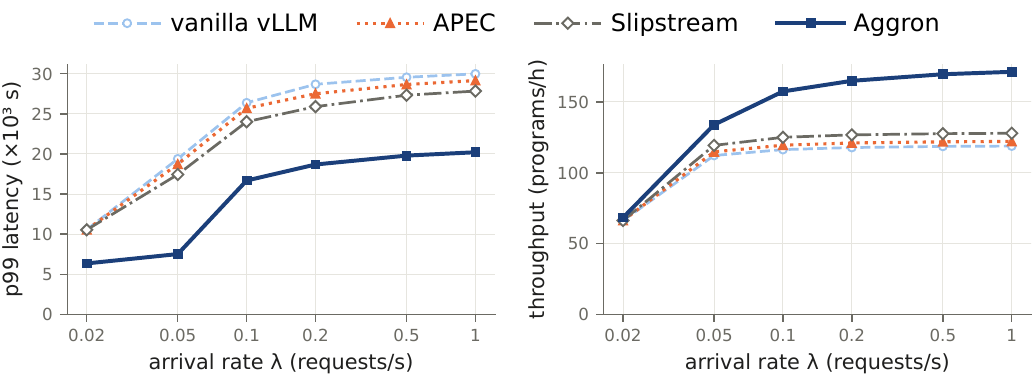}
        \caption{}
        \label{fig:tpt_p99}
    \end{subfigure}\hfill
    \begin{subfigure}[b]{0.5\linewidth}
        \centering
        \includegraphics[width=1\linewidth]{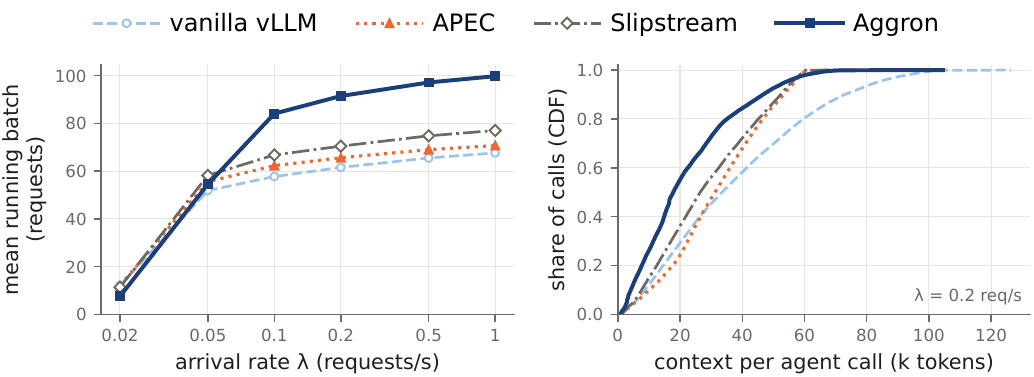}
        \caption{}
        \label{fig:compression_comp}
    \end{subfigure}
  \begin{subfigure}[b]{0.5\linewidth}
    \centering
    \includegraphics[width=1\linewidth]{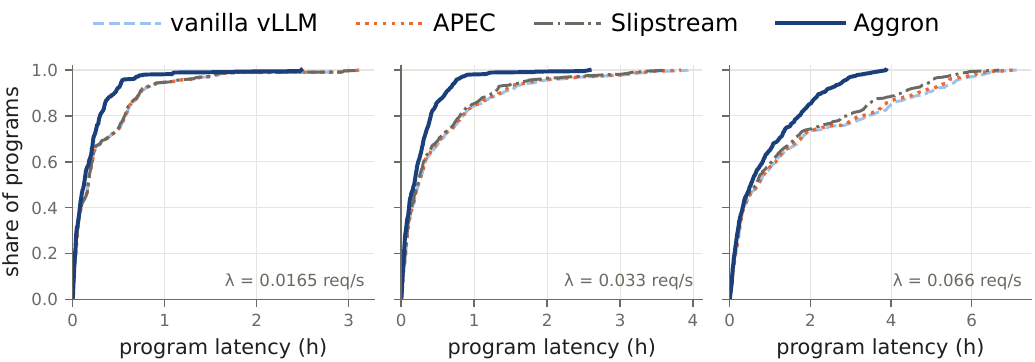}
    \caption{}
    \label{fig:latency_cdf}
  \end{subfigure}\hfill
  \begin{subfigure}[b]{0.5\linewidth}
    \centering
    \includegraphics[width=1\linewidth]{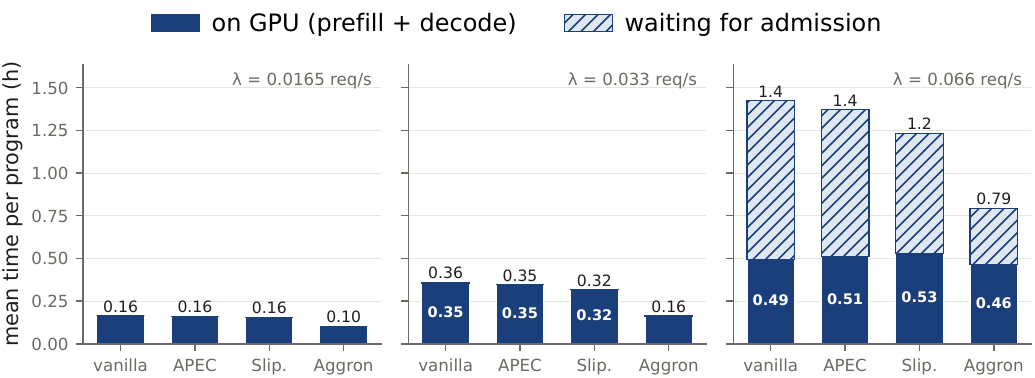}
    \caption{}
    \label{fig:latency_bkdn}
  \end{subfigure}
  \begin{subfigure}[b]{0.5\linewidth}
    \includegraphics[width=1\linewidth]{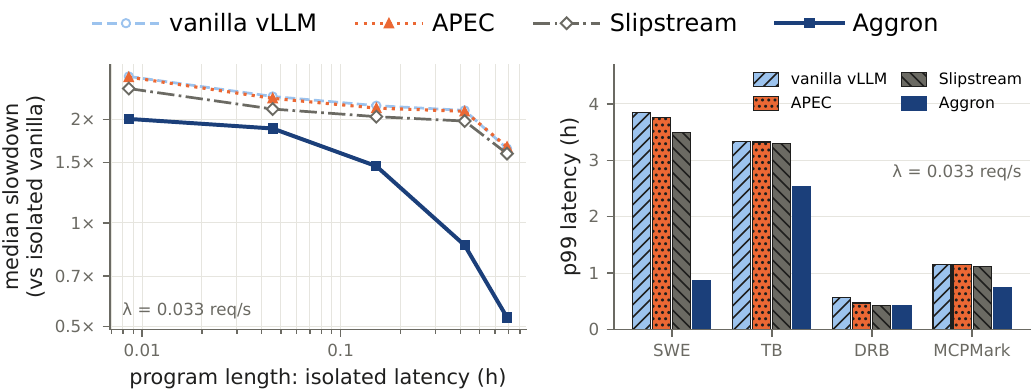}
    \caption{   }
    \label{fig:fairness}
  \end{subfigure}\hfill
  \begin{subfigure}[b]{0.5\linewidth}
    \centering
    \includegraphics[width=1\linewidth]{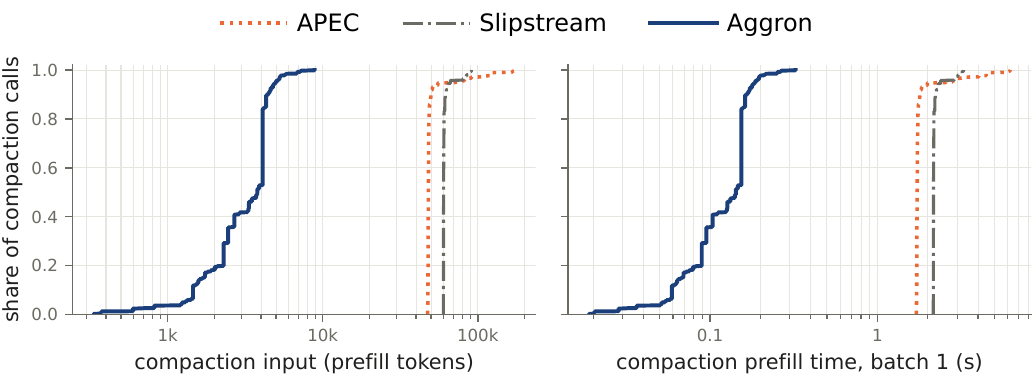}
    \caption{}
    \label{fig:summary_prefill}
  \end{subfigure}
  \vspace{-7mm}
  \caption{(a) P99 latency and throughput on Qwen3.6-35B-A3B. (b) Batch occupancy at
different rates, and the distribution of context size per inference call. (c) Latency
distribution and (d) latency breakdown at $0.35\times$, $0.7\times$ and $1.4\times$ the
saturation rate ($\approx 0.047$\,req/s). (e) Slowdown against program length, normalized
to isolated runs, and per-benchmark P99 latency, at $0.7\times$ the saturation rate.
(f) Compression statistics.}
  
  \Description{P99 Latency, and throughput evaluated on the Qwen3.6 model.Batch occupancy at different rates and distribution of context size when making an inference call. Latency distribution at different arrival rates. Latency breakdown at different arrival rates. Slowdown against request duration normalized to isolated runs, and benchmark-specific P99 latency at $\lambda$ near saturation. Compression statistics}
  \vspace{-3mm}
\end{figure*}

%% file: sections/6.eval.tex
\subsection{Performance analysis against SOTA}\label{sec:tpt_and_occ}
Figure~\ref{fig:tpt_p99} plots the P99 latency and throughput for requests arriving at different rates drawn from a Poisson distribution on the Qwen model. We compare our results to SOTA asynchronous summarization techniques Slipstream~\cite{slipstream2026} and APEC~\cite{apec2027}. At high loads ($\lambda = 1$), Aggron sustains $1.44\times$ the throughput of vanilla, $1.40\times$ APEC, and $1.34\times$ Slipstream, and cuts P99 latency by $33\%$, $31\%$ and $27\%$ respectively. This is because its smaller per-request KV footprint raises batch occupancy from $26\%$ to $39\%$ of the $256$-sequence limit, as seen in Figure~\ref{fig:compression_comp}. It also shows the distribution of the context size when an inference is made. This captures the variation in context size over the duration of the run and the summarization achieved across the methods. We see that Aggron has a larger share at smaller sizes, as the summarization of Aggron operates at smaller sizes. Once the thresholds for Slipstream and APEC are reached ($60$-$90$K), those methods start compressing, and their distributions catch up with Aggron's.

To understand the trend in P99, Figure~\ref{fig:latency_cdf} plots the distribution of the request latency. At lower rates, when inference dominates the latency, we get similar latencies across the methods. As the rate increases, the batch size increases by $3\times$, and inference gets slower. At higher rates, when memory saturates, the queue dominates latency. Figure~\ref{fig:latency_bkdn} plots the latency breakdown. Here Aggron's benefit comes from two factors. First, the improved batch occupancy drains the queue faster, so every request spends less time waiting. Second, agents do not contend for resources with summary requests. We examine this further in Section~\ref{sec:char_aggron}.

Aggron's gain grows with program length; we examine this in Figure~\ref{fig:fairness}. In the longest bin, its median slowdown is $0.53\times$, against $1.64\times$ for vanilla; in the middle bin it is $1.46\times$ against $2.19\times$. Slowdown here is measured against isolated vanilla. Each turn a program takes is cheaper to prefill and decode and frees KV sooner, so the benefit accumulates over long programs. The freed memory raises batch occupancy, which shortens the admission queue for every program. Short programs rarely compact, yet their median slowdown still falls from $2.67\times$ to $2.01\times$, because they wait in the same admission queue. Figure~\ref{fig:fairness} also shows the per-benchmark P99 latency. SWE benefits most ($4.5\times$ lower P99 than vanilla) as it spends the largest share of its time in the model ($59\%$) and the cap removes the most context ($74\%$). TB and MCPMark spend most of their time in tool execution, which summarization cannot shorten, so they gain only $1.3\times$ and $1.6\times$. DRB is tool-bound as well, and Aggron matches Slipstream there ($0.42$\,h).

From the discussion so far, Aggron's gain comes from its smaller per-turn context; Figure~\ref{fig:summary_prefill} shows that Aggron issues summarization at smaller thresholds. The smaller size of summarizations produces faster inferences and demands a smaller share of the KV cache. Aggron's largest summary is still about $5\times$ smaller than the smallest async summarization. At the median, the gap is $13$-$16\times$ in input size and $12$-$15\times$ in prefill time. So Aggron makes about $3\times$ more summarization calls, but each one is small enough to hide in queuing delay. The async systems make few, large ones. Asynchronous methods can also operate at Aggron's thresholds but with accuracy degradation, as they compress prose and non-prose alike (Table~\ref{tab:ctx-accuracy}, $4$K). Even if they compress only prose, an asynchronous compactor lets the agent continue when the summary has not landed, which Figure~\ref{fig:Queue share} shows lengthens head-of-line stalls. A synchronous $4$K prose cap reaches the same throughput on these traces, since each $4$K summary is small next to the queue wait (Figure~\ref{fig:Queue share}); Aggron's queue-time issue keeps that true as thresholds fall below $4$K and summaries grow costlier on the frontier models of Section~\ref{sec:scaling}.

\input{figures/Eval/acc_table}
\subsection{Accuracy comparisons}\label{sec:acc}
Table~\ref{tab:accuracy} reports accuracy for the three arms across three models
and four benchmarks. Accuracy rises in some cells and falls in others, but every
difference lies within the run-to-run noise band, so both compressed arms are as
accurate as vanilla. Since grow-4k and grow-1k sit at the two ends of Aggron's
threshold range, we expect Aggron's accuracy to lie between them.

The noise band varies widely across benchmarks, and is widest on
Terminal-Bench ($\pm 12$-$20$\,pp, against $\pm 7$-$9$\,pp on SWE-bench). The band
shrinks with the number of tasks and grows with the number on which two
identical runs disagree, and Terminal-Bench is worst on both. It has the fewest
tasks ($30$-$40$ scored), and many of its tasks are long, open-ended shell sessions
whose outcome flips between identical runs. A single early mistake, such as a
failed install or a wrong path, can derail the rest of the session. Its
differences are therefore the least resolved in the table, not evidence of a
larger effect.

Summarization also leaves trajectories within run-to-run variation in $22$ of $24$
arm-benchmark pairs. For each pair, a sign test counts the tasks on which summarization
moves the trajectory further from vanilla than a second vanilla run does, and in $22$
pairs that count is not significant once corrected for running $24$ tests, and the Jaccard excess is at
most $0.08$ in every cell. The two exceptions are SWE grow-1k on Qwen
($\Delta d = +0.11$) and GLM ($+0.07$). These agents make $1.3$-$1.7\times$ as
many tool calls as vanilla because they re-check their state after each summary.
The extra calls are insertions, which raise $\Delta d$, while the order of tools
is unchanged. The largest raw excess, DRB grow-4k on GLM ($+0.16$) and Gemma
($+0.11$), is not significant. DRB's tool names include the domain each search
fetches, so citing a different source counts as a substitution, and its null is
itself wide.


\subsection{Ablation studies}\label{sec:char_aggron}
\begin{figure*}[!t]
  \centering
  \captionsetup[sub]{skip=2pt}
  \begin{minipage}[t]{0.495\linewidth}
    \vspace{0pt}
    \begin{subfigure}{\linewidth}
      \includegraphics[width=\linewidth,trim=0 7 7 7,clip]{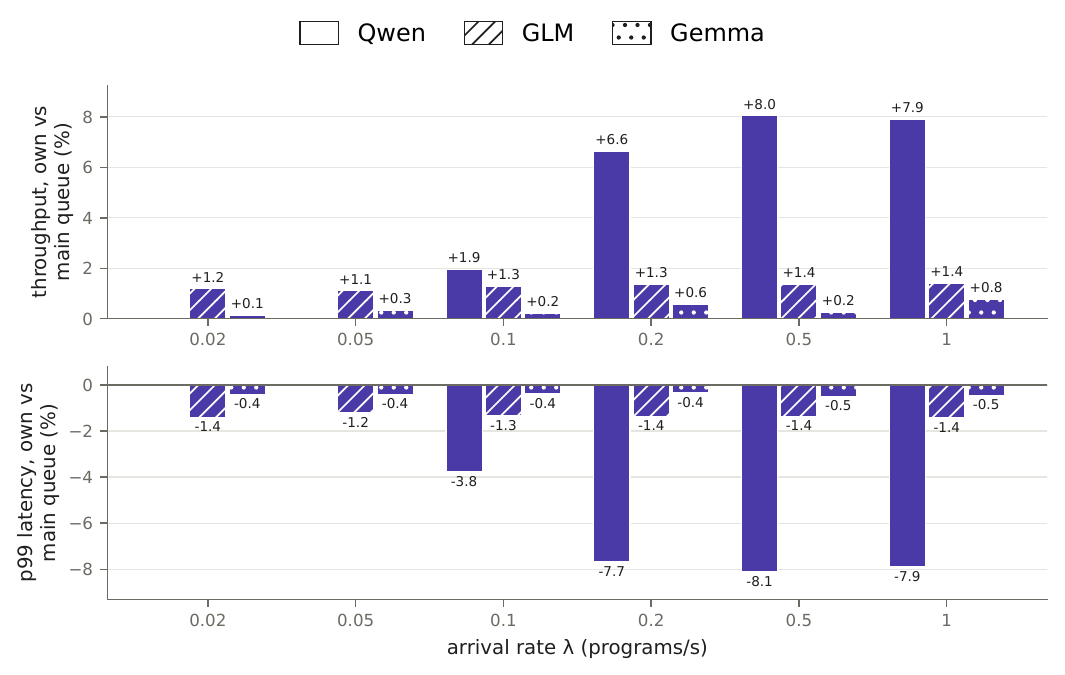}
      \caption{}
      \label{fig:split_q_p99_tpt}
    \end{subfigure}
    \setcounter{subfigure}{2}
    \begin{subfigure}{\linewidth}
      \includegraphics[width=\linewidth,trim=7 7 7 7,clip]{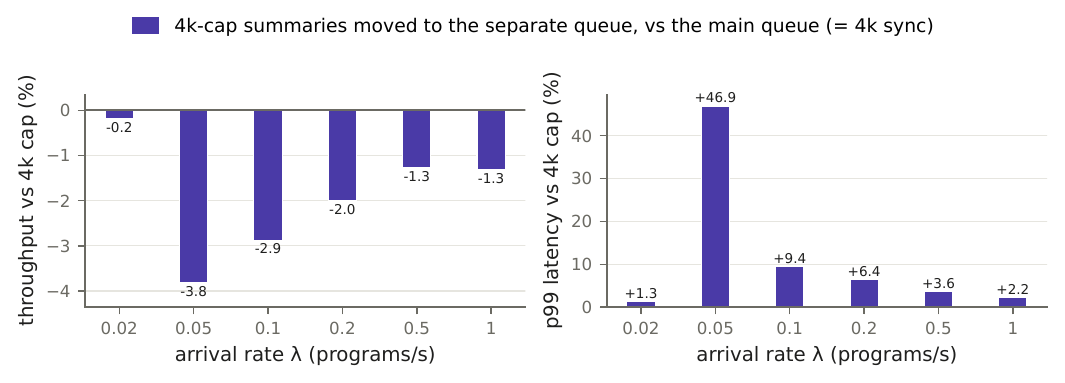}
      \caption{}
      \label{fig:split_q_cap}
    \end{subfigure}
  \end{minipage}\hfill
  \begin{minipage}[t]{0.495\linewidth}
    \vspace{0pt}
    \setcounter{subfigure}{1}
    \begin{subfigure}{\linewidth}
      \includegraphics[width=\linewidth,trim=7 7 7 7,clip]{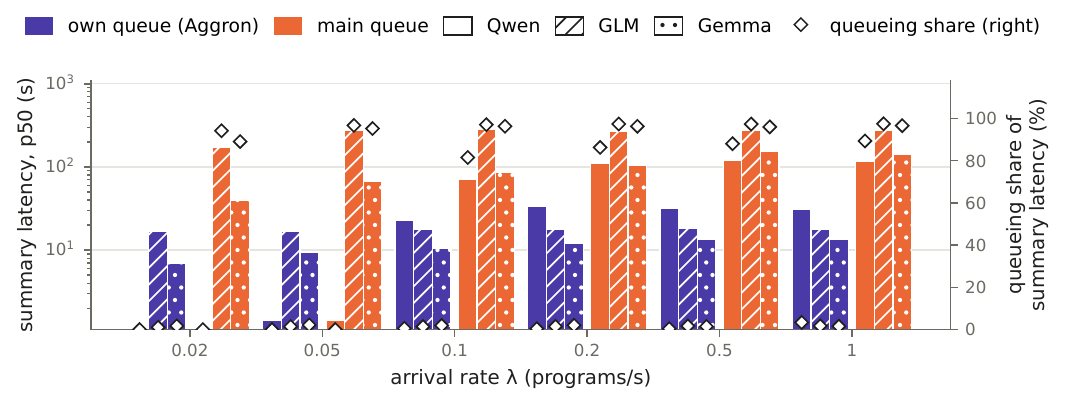}
      \caption{}
      \label{fig:split_q_latency}
    \end{subfigure}
    \setcounter{subfigure}{3}
    \begin{subfigure}{\linewidth}
      \includegraphics[width=\linewidth,trim=7 7 7 7,clip]{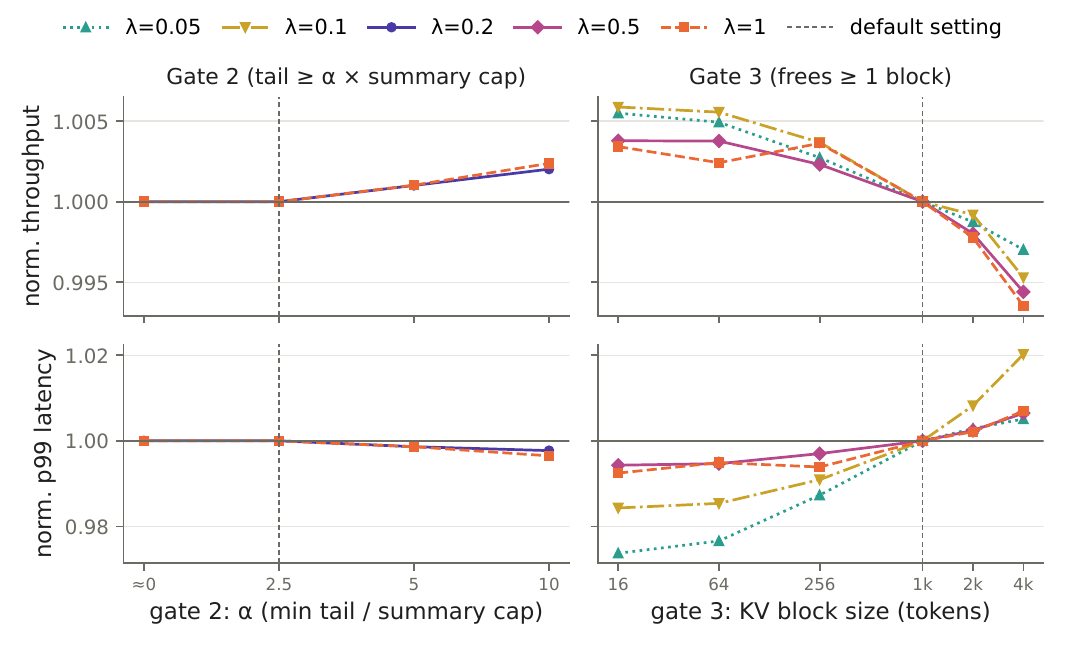}
      \caption{}
      \label{fig:gate_ablation}
    \end{subfigure}
  \end{minipage}
  \vspace{-1mm}
  \caption{(a) Throughput and P99 of the separate vs.\ shared summary queue.
  (b) P50 summary latency and queuing share of summary latency.
  (c) Throughput and P99 when the $4$K cap's own summaries move to the separate queue, against keeping them in the main queue.
  (d) Gate ablations, normalized to the default setting. }
  \Description{P99 and throughput of the separate vs. shared summary queue. P50 summary latency and queuing share of summary latency. P99 and throughput when the 4K cap's summaries move to the separate queue. Gate ablations.}
  \vspace{-3mm}
\end{figure*}

We first isolate the queue design, comparing Aggron's separate summary queue against a shared
queue in which summaries wait in the main queue alongside agent calls
(Figure~\ref{fig:split_q_p99_tpt}). Both fire the same tail-only summaries, so only the summary waits differ. Once
summaries fire ($\lambda \geq 0.1$), the separate queue raises throughput by $1.9$-$8.0\%$ and lowers
P99 by $3.8$-$8.1\%$ on Qwen, by $1.1$-$1.4\%$ and $1.2$-$1.4\%$ on GLM, and by $0.1$-$0.8\%$ and
$0.4$-$0.5\%$ on Gemma. Each stalled summary blocks the head for $8$-$14$\,s on all three
models, but Qwen fits batches of $68$-$77$ requests and finishes in a quarter to a third of the
time GLM and Gemma take at batches of $11$-$19$, so the same stalls fill more of its run.
The difference lies in summary latency (Figure~\ref{fig:split_q_latency}). In the shared queue, a summary waits
behind the agent calls ahead of it. Head-of-line stalls make up $82$-$98\%$ of its latency; its median latency reaches
$40$-$280$\,s, and none land before their call reaches the head of the queue, which then waits on it
for $8$-$40\%$ of the run. Aggron admits its summaries almost immediately, so they land in
$7$-$33$\,s and $93$-$100\%$ arrive in time. When summaries can be hidden, a separate queuing strategy improves performance.

Next, we measure the cost of raising summarization requests that cannot be hidden. Since we force the agentic request to stall at the head when its summary is in flight, performance can degrade if the summary does not complete before the agentic request is served. This can arise due to insufficient capacity for summaries to run, stalling the summarization queue. Figure~\ref{fig:split_q_cap} masks the controller and moves
the $4$K cap's own summarizations, with no early summaries, from the main queue into the separate one. The head then waits on its own summary for $12$-$49\%$ of the run, so the separate queue
loses $0.2$-$3.8\%$ throughput and adds $1.3$-$47\%$ to P99 relative to keeping the cap synchronous.
The loss peaks at light load ($\lambda = 0.05$), where the queue is too short to hide a summary. This is why Aggron raises a summary only when gate~1 predicts it can be hidden. On the frontier models of Section~\ref{sec:scaling}, gate~1 passes $89$-$99\%$ of candidate summaries, over $99.5\%$ of them land in time, and the head waits on its own summary for under $0.3\%$ of the run.

Predicting if the summarization can be hidden is the first gate of the controller. The other two gates check if the summarization is fruitful. Gate~2 requires the
tail to be at least $\alpha$ times the summary cap, and gate~3 requires the summary to free
at least one KV block. Figure~\ref{fig:gate_ablation} sweeps both, normalized to
the default setting ($\alpha = 2.5$, $1056$-token blocks). Summaries are typically
$\sim 10\times$ smaller than their tails, so neither gate changes which summaries
fire at the default setting. Removing gate~2 ($\alpha \approx 0$) leaves throughput and
P99 unchanged, and gate~3 declines at most one check. Tightening gate~2 to $\alpha = 10$
drops $63$-$66\%$ of summaries but moves throughput and P99 by under $0.4\%$.
Gate~3 starts declining only at $4$K-token blocks, where one block is as large as
the tail. Across $16$-token to $4$K-token blocks, throughput moves by at most $0.65\%$
and P99 by at most $2.6\%$, mainly because block size sets allocation granularity
for every request.

\subsection{Scaling Aggron}\label{sec:scaling}
\begin{figure*}[t]
  \centering
  \includegraphics[width=0.55\linewidth]{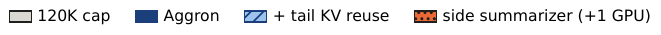}\\[2pt]
  \begin{minipage}[t]{0.32\linewidth}
    \centering
    \includegraphics[width=0.9\linewidth]{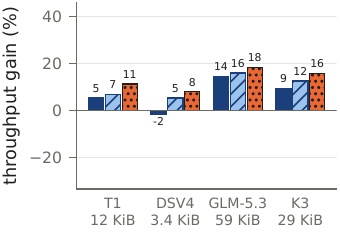}\\
    {\small (a) Throughput gain, $120$K cap}
  \end{minipage}\hfill
  \begin{minipage}[t]{0.32\linewidth}
    \centering
    \includegraphics[width=0.9\linewidth]{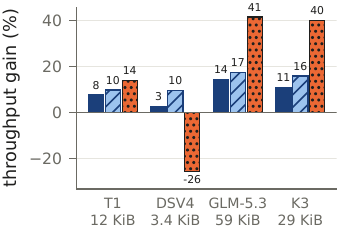}\\
    {\small (b) Throughput gain, $24$K cap}
  \end{minipage}\hfill
  \begin{minipage}[t]{0.32\linewidth}
    \centering
    \includegraphics[width=0.9\linewidth]{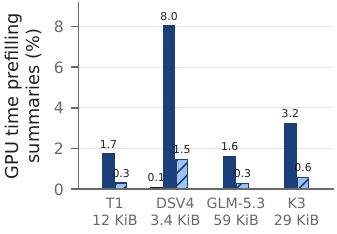}\\
    {\small (c) Summary prefill, $120$K cap}
  \end{minipage}
  \vspace{-2mm}
  \caption{Aggron on four frontier models (shared trace mix, $\lambda = 1$). (a) and (b)
compare each arm with the cap it runs under. The side summarizer uses one more GPU.
Labels under each model give its KV per token per GPU.}
  \label{fig:sec4}
  \Description{Aggron on four frontier models (shared trace mix, $\lambda = 1$). (a) and (b)
compare each arm with the cap it runs under. The side summarizer uses one more GPU.
Labels under each model give its KV per token per GPU.}
\vspace{-3mm}
\end{figure*}

Under the $120$K cap (Figure \ref{fig:sec4}a), Aggron raises throughput by $5$-$14\%$ on T1, K3 and GLM-5.3 and
cuts P99 latency by $7$-$13\%$. Its summaries cut cap restarts from $1{,}468$ to at most $226$, and the memory they free
admits more requests. The gain grows with KV per token, from $+5\%$ on T1 ($12$\,KiB) to
$+14\%$ on GLM-5.3 ($59$\,KiB). DSV4 ($3.4$\,KiB) fills only $68\%$ of HBM uncompressed,
so freed memory buys nothing while summary prefill takes $8\%$ of serving-GPU time
($1.6$-$3.2\%$ elsewhere), and it loses $1.7\%$. If each summary reuses the cached KV of the prose it compresses and recomputes $15\%$
of it~\cite{cacheblend}, this share falls to $0.3$-$1.5\%$, and throughput rises on all four models by $5$-$16\%$, with DSV4 at $+5.3\%$, and P99 falls by $5$-$14\%$.

Under the $24$K cap (Figure \ref{fig:sec4}b), Aggron halves the baseline's $49{,}421$ restarts, which offsets its
summary prefill, and
raises throughput by $3$-$14\%$ and cuts P99 by $4$-$12\%$, or $10$-$17\%$ and
$10$-$15\%$ with tail reuse.
Moving every summary to a 35B side model frees the serving GPU but cannot reuse its
KV. At $120$K it gains $8$-$18\%$, only $2$-$4$ points more than tail reuse, with a
second GPU. At $24$K the side GPU runs $83\%$ busy. It prefills about $13$K tokens/s,
faster than GLM-5.3 and K3 ($+40$-$41\%$) but a third of DSV4's, where it loses $26\%$
as requests wait on their own summary for $81\%$ of the run. Routing each summarization request dynamically to the less
loaded of the two is future work. Gains trail Section~\ref{sec:tpt_and_occ}, where Qwen's programs fit its $131$K window and vanilla never compacts. Here most programs outgrow the $120$K cap (median peak $163$K), so the baseline compacts as much as Aggron or more.

On four frontier models, Aggron raises throughput by up to $14\%$ over either a $120$K or a
$24$K cap, and by $5$-$17\%$ on all four when summaries reuse the cached KV of the prose they replace.

%% file: figures/Eval/acc_table.tex
\begin{table*}[t]
\centering
\small
\caption{Task accuracy and trajectory divergence under compaction
(Section~\ref{sec:setup}).}
\vspace{-3mm}
\label{tab:accuracy}
\begin{tabular}{|llrrrrrcrr|}
\toprule
Model & Benchmark & grow-1k & grow-4k & vanilla & $\Delta_\text{1k}$ & $\Delta_\text{4k}$ & Noise ($\pm$) & $\Delta d_\text{1k}$ & $\Delta d_\text{4k}$ \\
\midrule
\multirow{4}{*}{Qwen3.6-35B-A3B}
  & SWE    & 67.7  & 66.3  & 68.1  & $-1.1$  & $-1.1$  & 6.7  & $+0.11$ & $0.00$ \\
  & TerminalBench           & 55.0 & 57.5 & 50.0 & $+5.0$ & $+7.5$ & 12.0 & $0.00$ & $+0.04$ \\
  & MCPMark & 47.0  & 42.4  & 39.4  & $+7.6$  & $+3.0$  & 7.9  & $+0.08$ & $0.00$ \\
  & DeepResearchBench     & 42.91 & 43.46 & 43.55 & $-0.72$ & $-0.10$ & 1.9  & $+0.03$ & $0.00$ \\
\midrule
\multirow{4}{*}{GLM-4.7-Flash}
  & SWE        & 55.0 & 57.5 & 56.2 & $-1.2$ & $+1.2$ & 9.2 & $+0.07$ & $+0.04$ \\
  & TerminalBench         & 26.7  & 36.7  & 31.7  & $-5.0$  & $+5.0$  & 19.6 & $+0.01$ & $0.00$ \\
  & MCPMark    & 34.9  & 34.9  & 35.7  & $-0.8$  & $-0.8$  & 8.2  & $-0.04$ & $-0.04$ \\
  & DeepResearchBench        & 41.16 & 42.27 & 41.12 & $+0.04$ & $+1.15$ & 1.9  & $+0.05$ & $+0.16$ \\
\midrule
\multirow{4}{*}{Gemma-4-26B-A4B}
  & SWE        & 62.7 & 62.7 & 65.3 & $-2.7$ & $-2.7$ & 7.4 & $0.00$ & $0.00$ \\
  & TerminalBench         & 39.4  & 48.5  & 39.4  & $0.0$   & $+9.1$  & 16.8 & $-0.02$ & $0.00$ \\
  & MCPMark    & 45.5  & 45.5  & 40.2  & $+5.3$  & $+5.3$  & 6.6  & $0.00$ & $0.00$ \\
  & DeepResearchBench        & 42.35 & 40.63 & 40.55 & $+1.80$ & $+0.08$ & 3.0  & $+0.07$ & $+0.11$ \\
\bottomrule
\end{tabular}

\end{table*}

%% file: sections/7.related.tex
\textbf{Context summarization.}
Most work on agent context summarization either trains the model to summarize
better or decides \emph{when} to summarize. ACON~\cite{acon2025} optimizes the
summarization guideline from failed trajectories; ReSum~\cite{resum2025},
MEM1~\cite{mem12025}, AgentFold~\cite{agentfold2026} and
CompactionRL~\cite{compactionrl2026} train agents to fold or rewrite their own
history with supervised or reinforcement learning; SelfCompact~\cite{selfcompact2026}
lets the model choose when to compact, and MemGPT~\cite{memgpt2023} pages context
between the window and external memory. These methods summarize the whole
trajectory and rewrite it in place, typically near the context limit, and none
of them considers the serving system. The Complexity Trap~\cite{complexitytrap2025}
shows that simply masking old observations rivals LLM summarization. The agents in
our traces, however, revisit old tool outputs (Section~\ref{sec:motivation}), so
Aggron keeps observations and compresses the prose around them.
Belcak et al.~\cite{belcak2026longcontext} ask how much context an agent needs,
and show that frontier agents that compact their full history at budgets as
small as $16$K tokens and restart from the summary lose little to no accuracy on
SWE-bench and Terminal-Bench. Their agent compacts in-band at a fixed threshold
and re-prefills the restarted context, and serving cost is measured per token on
isolated fixed-length requests. Aggron instead asks how to serve summarization on a
loaded, shared server. The server decides when to compress from queue state,
summarizes only prose while the request waits for admission, keeps tool outputs
verbatim, and freezes each summary into an append-only prefix so the cached KV
survives.

\textbf{KV-cache eviction.}
Eviction drops the tokens the model is predicted not to need
(H2O~\cite{zhang2023h2o}, Scissorhands~\cite{liu2023scissorhands},
StreamingLLM~\cite{xiao2024streamingllm}, SnapKV~\cite{snapkv2024},
PyramidInfer~\cite{yang2024pyramidinfer}), and recent work adapts eviction and
attention matching to agents~\cite{agentkv2025,onlinekvcompaction2026,attnmatching2026}.
Eviction is lossy for every token, including tool outputs, and its memory savings
do not reliably translate into serving throughput~\cite{kvpitfalls2025}. It is
orthogonal to Aggron and can be combined with it.

\textbf{Asynchronous summarization.}
Several systems move summarization off the agent's critical path.
APEC~\cite{apec2027} summarizes a snapshot in the background once the context
passes a soft threshold and installs it at the next request;
Slipstream~\cite{slipstream2026} runs the compactor beside the agent and
validates the summary with a judge; CoMem~\cite{comem2026} overlaps a trained
memory model with agent inference; Parallel Context
Compaction~\cite{parallelcompaction2026} splits one compaction across parallel
requests; and sleep-time compute~\cite{sleeptime2025} uses user idle time. All
of them hide compaction behind a single agent's own execution and are evaluated
on a dedicated server. Under load, their summaries compete with other requests
for the same KV and batch slots. Aggron instead hides compaction inside the
queuing delay that load creates, and runs it only in capacity the waiting
requests cannot use.

\textbf{Serving schedulers.}
LLM serving builds on continuous batching~\cite{yu2022orca}, paged KV
memory~\cite{kwon2023pagedattention} and chunked prefill~\cite{agrawal2023sarathi}.
Schedulers fight head-of-line blocking by preemption~\cite{fastserve2023},
learned shortest-job-first ordering~\cite{ltr2024}, or queue
reordering and eviction~\cite{qlm2024}. Agent-aware systems retain or move KV
across tool calls. InferCept~\cite{infercept2024} preserves, discards or swaps
it; Continuum~\cite{continuum2024} pins it with a time-to-live; MORI~\cite{MORI}
offloads idle programs; and Autellix~\cite{autellix2025},
Pythia~\cite{PYTHIA} and ThunderAgent~\cite{thunderagent2026} schedule at the
program level. These systems reorder, move or retain KV, but none reduces the
KV a waiting request will need. Aggron keeps vLLM's FCFS order and shrinks the
waiting requests themselves.

%% file: sections/8.Conclusion.tex
To maximize the cost-effectiveness of summarization, this paper presented a novel
serving
system called {\em Aggron}. Compared to the Slipstream [10] and
APEC [36] state-of-the-art asynchronous summarization schemes,
Aggron reduces P99 latency by 27\% and 31\%, reduces mean latency
by 11\% and 15\%, and increases throughput by 1.34× and 1.40×,
respectively. Aggron’s accuracy stays
within run-to-run noise of uncompressed execution. On four frontier models, it raises throughput by up to $14\%$ over
either a $120$K or a $24$K cap, and by $5$-$17\%$ on all four when summaries reuse
the cached KV of the prose they replace.



%% file: refs.bib
@inproceedings{yu2022orca,
  author    = {Yu, Gyeong-In and Jeong, Joo Seong and Kim, Geon-Woo and
               Kim, Soojeong and Chun, Byung-Gon},
  title     = {Orca: A Distributed Serving System for Transformer-Based
               Generative Models},
  booktitle = {16th USENIX Symposium on Operating Systems Design and Implementation (OSDI 22)},
  series    = {OSDI '22},
  pages     = {521--538},
  year      = {2022},
  publisher = {USENIX Association},
  address   = {Berkeley, CA, USA},
  url       = {https://www.usenix.org/conference/osdi22/presentation/yu}
}

@inproceedings{kwon2023pagedattention,
  author    = {Kwon, Woosuk and Li, Zhuohan and Zhuang, Siyuan and Sheng,
               Ying and Zheng, Lianmin and Yu, Cody Hao and Gonzalez, Joseph
               and Zhang, Hao and Stoica, Ion},
  title     = {Efficient Memory Management for Large Language Model Serving
               with {PagedAttention}},
  booktitle = {Proceedings of the 29th Symposium on Operating Systems Principles},
  series    = {SOSP '23},
  pages     = {611--626},
  year      = {2023},
  publisher = {Association for Computing Machinery},
  address   = {New York, NY, USA},
  doi       = {10.1145/3600006.3613165},
  url       = {https://dl.acm.org/doi/10.1145/3600006.3613165}
}

@misc{continuum2024,
  author       = {Li, Hanchen and
               He, Runyuan and
               Mang, Qiuyang and
               Zhang, Qizheng and
               Mao, Huanzhi and
               Chen, Xiaokun and
               Zhou, Hangrui and
               Zhang, Huanchen and
               Cheung, Alvin and
               Gonzalez, Joseph and
               Stoica, Ion},
  title        = {Continuum: Efficient and Robust Multi-Turn {LLM} Agent
                   Scheduling with {KV} Cache Time-to-Live},
  year         = {2026},
  howpublished = {In \emph{ICLR 2026 Workshop on Lifelong Agents: Learning, Aligning, Evolving}},
  url          = {https://openreview.net/forum?id=sVzK0LC9pn}
}

@inproceedings{yang2024sweagent,
  author    = {Yang, John and Jimenez, Carlos E. and Wettig, Alexander and
                   Lieret, Kilian and Yao, Shunyu and Narasimhan, Karthik and
                   Press, Ofir},
  title     = {{SWE}-agent: Agent-Computer Interfaces Enable Automated
                   Software Engineering},
  booktitle = {Advances in Neural Information Processing Systems},
  series    = {NeurIPS '24},
  volume    = {37},
  pages     = {50528--50652},
  year      = {2024},
  publisher = {Curran Associates, Inc.},
  address   = {Red Hook, NY, USA},
  doi       = {10.52202/079017-1601},
  url       = {https://proceedings.neurips.cc/paper_files/paper/2024/hash/5a7c947568c1b1328ccc5230172e1e7c-Abstract-Conference.html}
}

@inproceedings{yao2022react,
  author    = {Yao, Shunyu and Zhao, Jeffrey and Yu, Dian and Du, Nan and
                   Shafran, Izhak and Narasimhan, Karthik and Cao, Yuan},
  title     = {{ReAct}: Synergizing Reasoning and Acting in Language Models},
  booktitle = {International Conference on Learning Representations},
  series    = {ICLR '23},
  numpages  = {33},
  year      = {2023},
  publisher = {OpenReview.net},
  address   = {Kigali, Rwanda},
  url       = {https://openreview.net/forum?id=WE_vluYUL-X}
}

@inproceedings{swebench2023,
  author    = {Jimenez, Carlos E. and Yang, John and Wettig, Alexander and
                   Yao, Shunyu and Pei, Kexin and Press, Ofir and
                   Narasimhan, Karthik},
  title     = {{SWE}-bench: Can Language Models Resolve Real-World {GitHub}
                   Issues?},
  booktitle = {The Twelfth International Conference on Learning Representations},
  series    = {ICLR '24},
  pages     = {54107--54157},
  year      = {2024},
  publisher = {OpenReview.net},
  address   = {Vienna, Austria},
  url       = {https://proceedings.iclr.cc/paper_files/paper/2024/hash/edac78c3e300629acfe6cbe9ca88fb84-Abstract-Conference.html}
}

@inproceedings{ainslie2023gqa,
  author    = {Ainslie, Joshua and Lee-Thorp, James and de Jong, Michiel and
                   Zemlyanskiy, Yury and Lebr{\'o}n, Federico and Sanghai, Sumit},
  title     = {{GQA}: Training Generalized Multi-Query Transformer Models
                   from Multi-Head Checkpoints},
  booktitle = {Proceedings of the 2023 Conference on Empirical Methods in Natural Language Processing},
  series    = {EMNLP '23},
  pages     = {4895--4901},
  year      = {2023},
  publisher = {Association for Computational Linguistics},
  address   = {Stroudsburg, PA, USA},
  doi       = {10.18653/v1/2023.emnlp-main.298},
  url       = {https://aclanthology.org/2023.emnlp-main.298/}
}

@inproceedings{vaswani2017attention,
  author    = {Vaswani, Ashish and Shazeer, Noam and Parmar, Niki and
               Uszkoreit, Jakob and Jones, Llion and Gomez, Aidan N. and
               Kaiser, {\L}ukasz and Polosukhin, Illia},
  title     = {Attention Is All You Need},
  booktitle = {Advances in Neural Information Processing Systems 30 (NeurIPS 2017)},
  series    = {NeurIPS '17},
  volume    = {30},
  pages     = {5998--6008},
  year      = {2017},
  publisher = {Curran Associates, Inc.},
  address   = {Red Hook, NY, USA},
  url       = {https://proceedings.neurips.cc/paper_files/paper/2017/hash/3f5ee243547dee91fbd053c1c4a845aa-Abstract.html}
}

@inproceedings{pope2023efficiently,
  author    = {Pope, Reiner and Douglas, Sholto and Chowdhery, Aakanksha and
               Devlin, Jacob and Bradbury, James and Heek, Jonathan and
               Xiao, Kefan and Agrawal, Shivani and Dean, Jeff},
  title     = {Efficiently Scaling Transformer Inference},
  booktitle = {Proceedings of Machine Learning and Systems},
  series    = {MLSys '23},
  volume    = {5},
  pages     = {606--624},
  year      = {2023},
  publisher = {mlsys.org},
  address   = {Miami, FL, USA},
  url       = {https://proceedings.mlsys.org/paper_files/paper/2023/hash/c4be71ab8d24cdfb45e3d06dbfca2780-Abstract-mlsys2023.html}
}

@inproceedings{sheng2023flexgen,
  author    = {Sheng, Ying and
               Zheng, Lianmin and
               Yuan, Binhang and
               Li, Zhuohan and
               Ryabinin, Max and
               Chen, Beidi and
               Liang, Percy and
               R{\'e}, Christopher and
               Stoica, Ion and
               Zhang, Ce},
  title     = {{FlexGen}: High-Throughput Generative Inference of Large
               Language Models with a Single {GPU}},
  booktitle = {Proceedings of the 40th International Conference on Machine Learning},
  series    = {ICML '23},
  volume    = {202},
  pages     = {31094--31116},
  year      = {2023},
  publisher = {PMLR},
  address   = {Honolulu, HI, USA},
  url       = {https://proceedings.mlr.press/v202/sheng23a.html}
}

@misc{agrawal2023sarathi,
  title         = {{SARATHI}: Efficient {LLM} Inference by Piggybacking Decodes
                   with Chunked Prefills},
  author        = {Agrawal, Amey and Panwar, Ashish and Mohan, Jayashree and
                   Kwatra, Nipun and Gulavani, Bhargav S. and Ramjee, Ramachandran},
  year          = {2023},
  eprint        = {2308.16369},
  archivePrefix = {arXiv}
}

@inproceedings{merrill2026terminal,
  author    = {Merrill, Mike A. and Shaw, Alexander G. and Carlini, Nicholas and
                   Li, Boxuan and Raj, Harsh and Bercovich, Ivan and Shi, Lin and
                   Shin, Jeong Yeon and Walshe, Thomas and Buchanan, E. Kelly and others},
  title     = {{Terminal-Bench}: Benchmarking Agents on Hard, Realistic Tasks in
                   Command Line Interfaces},
  booktitle = {The Fourteenth International Conference on Learning Representations},
  series    = {ICLR '26},
  pages     = {40903--40986},
  year      = {2026},
  publisher = {OpenReview.net},
  address   = {Rio de Janeiro, Brazil},
  url       = {https://proceedings.iclr.cc/paper_files/paper/2026/hash/444a3737adaee10d86ad2ef5f74468e6-Abstract-Conference.html}
}

@inproceedings{thunderagent2026,
  author    = {Kang, Hao and
               Li, Ziyang and
               Xu, Weili and
               Yang, Xinyu and
               Chen, Yinfang and
               Wang, Junxiong and
               Chen, Beidi and
               Krishna, Tushar and
               Xu, Chenfeng and
               Arora, Simran},
  title     = {{ThunderAgent}: A Fast, Simple, and {Program-Aware} Agentic Inference System},
  booktitle = {Proceedings of the 43rd International Conference on Machine Learning},
  series    = {ICML '26},
  volume    = {306},
  pages     = {55630--55657},
  year      = {2026},
  publisher = {PMLR},
  address   = {Seoul, South Korea},
  url       = {https://proceedings.mlr.press/v306/kang26j.html}
}

@misc{slipstream2026,
  author        = {Zhuofu Chen and Rui Pan and Yinwei Dai and Ravi Netravali},
  title         = {Slipstream: Trajectory-Grounded Compaction Validation for
                   Long-Horizon Agents},
  year          = {2026},
  eprint        = {2605.08580},
  archivePrefix = {arXiv},
  url           = {https://arxiv.org/abs/2605.08580}
}

@misc{complexitytrap2025,
  author       = {Tobias Lindenbauer and Igor Slinko and Ludwig Felder and Egor
                   Bogomolov and Yaroslav Zharov},
  title        = {The Complexity Trap: Simple Observation Masking Is as Efficient
                   as {LLM} Summarization for Agent Context Management},
  year         = {2025},
  howpublished = {In \emph{Fourth Workshop on Deep Learning for Code (DL4C) at NeurIPS 2025}},
  url          = {https://arxiv.org/abs/2508.21433}
}

@inproceedings{kvpitfalls2025,
  author    = {Alex Chen and Renato Geh and Aditya Grover and Guy Van den
                   Broeck and Daniel Israel},
  title     = {The Pitfalls of {KV} Cache Compression},
  booktitle = {Proceedings of the 64th Annual Meeting of the Association for Computational Linguistics (Volume 1: Long Papers)},
  series    = {ACL '26},
  volume    = {1},
  pages     = {41530--41553},
  year      = {2026},
  publisher = {Association for Computational Linguistics},
  address   = {Stroudsburg, PA, USA},
  doi       = {10.18653/v1/2026.acl-long.1926},
  url       = {https://aclanthology.org/2026.acl-long.1926/}
}

@inproceedings{tcrallm2023,
  author    = {Junyi Liu and Liangzhi Li and Tong Xiang and Bowen Wang and
                   Yiming Qian},
  title     = {{TCRA-LLM}: Token Compression Retrieval Augmented Large Language
                   Model for Inference Cost Reduction},
  booktitle = {Findings of the Association for Computational Linguistics: EMNLP 2023},
  series    = {Findings of EMNLP '23},
  pages     = {9796--9810},
  year      = {2023},
  publisher = {Association for Computational Linguistics},
  address   = {Stroudsburg, PA, USA},
  doi       = {10.18653/v1/2023.findings-emnlp.655},
  url       = {https://aclanthology.org/2023.findings-emnlp.655/}
}

@inproceedings{yang2024pyramidinfer,
  author    = {Yang, Dongjie and Han, XiaoDong and Gao, Yan and Hu, Yao and
                   Zhang, Shilin and Zhao, Hai},
  title     = {{PyramidInfer}: Pyramid {KV} Cache Compression for
                   High-throughput {LLM} Inference},
  booktitle = {Findings of the Association for Computational Linguistics: ACL 2024},
  series    = {Findings of ACL '24},
  pages     = {3258--3270},
  year      = {2024},
  publisher = {Association for Computational Linguistics},
  address   = {Stroudsburg, PA, USA},
  doi       = {10.18653/v1/2024.findings-acl.195},
  url       = {https://aclanthology.org/2024.findings-acl.195/}
}

@inproceedings{zhang2023h2o,
  author    = {Zhang, Zhenyu and Sheng, Ying and Zhou, Tianyi and
               Chen, Tianlong and Zheng, Lianmin and Cai, Ruisi and
               Song, Zhao and Tian, Yuandong and R{\'e}, Christopher and
               Barrett, Clark and Wang, Zhangyang and Chen, Beidi},
  title     = {{H2O}: Heavy-Hitter Oracle for Efficient Generative Inference
               of Large Language Models},
  booktitle = {Advances in Neural Information Processing Systems 36 (NeurIPS 2023)},
  series    = {NeurIPS '23},
  volume    = {36},
  pages     = {34661--34710},
  year      = {2023},
  publisher = {Curran Associates, Inc.},
  address   = {Red Hook, NY, USA},
  doi       = {10.52202/075280-1506},
  url       = {https://proceedings.neurips.cc/paper_files/paper/2023/hash/6ceefa7b15572587b78ecfcebb2827f8-Abstract-Conference.html}
}

@inproceedings{chen2024nacl,
  author    = {Chen, Yilong and Wang, Guoxia and Shang, Junyuan and
                   Cui, Shiyao and Zhang, Zhenyu and Liu, Tingwen and
                   Wang, Shuohuan and Sun, Yu and Yu, Dianhai and Wu, Hua},
  title     = {{NACL}: A General and Effective {KV} Cache Eviction Framework for {LLM} at Inference Time},
  booktitle = {Proceedings of the 62nd Annual Meeting of the Association for Computational Linguistics (Volume 1: Long Papers)},
  series    = {ACL '24},
  volume    = {1},
  pages     = {7913--7926},
  year      = {2024},
  publisher = {Association for Computational Linguistics},
  address   = {Stroudsburg, PA, USA},
  doi       = {10.18653/v1/2024.acl-long.428},
  url       = {https://aclanthology.org/2024.acl-long.428/}
}

@inproceedings{liu2023scissorhands,
  author    = {Liu, Zichang and Desai, Aditya and Liao, Fangshuo and
               Wang, Weitao and Xie, Victor and Xu, Zhaozhuo and
               Kyrillidis, Anastasios and Shrivastava, Anshumali},
  title     = {Scissorhands: Exploiting the Persistence of Importance
               Hypothesis for {LLM} {KV} Cache Compression at Test Time},
  booktitle = {Advances in Neural Information Processing Systems},
  series    = {NeurIPS '23},
  volume    = {36},
  pages     = {52342--52364},
  year      = {2023},
  publisher = {Curran Associates, Inc.},
  address   = {Red Hook, NY, USA},
  doi       = {10.52202/075280-2279},
  url       = {https://proceedings.neurips.cc/paper_files/paper/2023/hash/a452a7c6c463e4ae8fbdc614c6e983e6-Abstract-Conference.html}
}

@inproceedings{xiao2024streamingllm,
  author    = {Xiao, Guangxuan and Tian, Yuandong and Chen, Beidi and Han, Song and Lewis, Mike},
  title     = {Efficient Streaming Language Models with Attention Sinks},
  booktitle = {The Twelfth International Conference on Learning Representations (ICLR 2024)},
  series    = {ICLR '24},
  pages     = {21875--21895},
  year      = {2024},
  publisher = {OpenReview.net},
  address   = {Vienna, Austria},
  url       = {https://proceedings.iclr.cc/paper_files/paper/2024/hash/5e5fd18f863cbe6d8ae392a93fd271c9-Abstract-Conference.html}
}

@inproceedings{acon2025,
  author    = {Minki Kang and Wei-Ning Chen and Dongge Han and Huseyin A. Inan
                   and Lukas Wutschitz and Yanzhi Chen and Robert Sim and Saravan
                   Rajmohan},
  title     = {{ACON}: Optimizing Context Compression for Long-Horizon {LLM}
                   Agents},
  booktitle = {Proceedings of the 43rd International Conference on Machine Learning},
  series    = {ICML '26},
  volume    = {306},
  pages     = {55321--55375},
  year      = {2026},
  publisher = {PMLR},
  address   = {Seoul, South Korea},
  url       = {https://proceedings.mlr.press/v306/kang26b.html}
}

@inproceedings{retrac2026,
  author    = {Jialiang Zhu and Gongrui Zhang and Xiaolong Ma and Lin Xu and
                   Miaosen Zhang and Ruiqi Yang and Song Wang and Kai Qiu and
                   Zhirong Wu and Qi Dai and Ruichun Ma and Bei Liu and Yifan Yang
                   and Chong Luo and Zhengyuan Yang and Linjie Li and Lijuan Wang
                   and Weizhu Chen and Xin Geng and Baining Guo},
  title     = {{RE-TRAC}: Recursive Trajectory Compression for Deep Search
                   Agents},
  booktitle = {Proceedings of the 43rd International Conference on Machine Learning},
  series    = {ICML '26},
  volume    = {306},
  pages     = {167986--168007},
  year      = {2026},
  publisher = {PMLR},
  address   = {Seoul, South Korea},
  url       = {https://proceedings.mlr.press/v306/zhu26bd.html}
}

@misc{agentkv2025,
  author        = {Liu, Taowen Tony and Wong, Jeffrey T. H. and Xiao, Can and
                   Yang, Bowen and Chen, Hao Mark and Zhao, Yiren},
  title         = {{AgentKV}: Phase-Aware {KV} Eviction for Agentic {LLMs}},
  year          = {2026},
  eprint        = {2609.14872},
  archiveprefix = {arXiv},
  url           = {https://arxiv.org/abs/2609.14872}
}

@misc{lmcache,
  author        = {Liu, Yuhan and
                   Cheng, Yihua and
                   Yao, Jiayi and
                   An, Yuwei and
                   Chen, Xiaokun and
                   Feng, Shaoting and
                   Huang, Yuyang and
                   Shen, Samuel and
                   Zhang, Rui and
                   Du, Kuntai and
                   Jiang, Junchen},
  title         = {{LMCache}: An Efficient {KV} Cache Layer for Enterprise-Scale
                   {LLM} Inference},
  year          = {2025},
  eprint        = {2510.09665},
  archivePrefix = {arXiv},
  url           = {https://arxiv.org/abs/2510.09665},
  doi           = {10.48550/arXiv.2510.09665}
}

@inproceedings{zheng2024sglang,
  author    = {Zheng, Lianmin and Yin, Liangsheng and Xie, Zhiqiang and
                   Sun, Chuyue and Huang, Jeff and Yu, Cody Hao and Cao, Shiyi
                   and Kozyrakis, Christos and Stoica, Ion and
                   Gonzalez, Joseph E. and Barrett, Clark and Sheng, Ying},
  title     = {{SGLang}: Efficient Execution of Structured Language Model
                   Programs},
  booktitle = {Advances in Neural Information Processing Systems},
  series    = {NeurIPS '24},
  volume    = {37},
  pages     = {62557--62583},
  year      = {2024},
  publisher = {Curran Associates, Inc.},
  address   = {Red Hook, NY, USA},
  doi       = {10.52202/079017-2000},
  url       = {https://proceedings.neurips.cc/paper_files/paper/2024/hash/724be4472168f31ba1c9ac630f15dec8-Abstract-Conference.html}
}

@inproceedings{du2025deepresearch,
  author    = {Du, Mingxuan and
               Xu, Benfeng and
               Zhu, Chiwei and
               Zhang, Licheng and
               Wang, Xiaorui and
               Mao, Zhendong},
  title     = {{DeepResearch Bench}: A Comprehensive Benchmark for Deep
                   Research Agents},
  booktitle = {International Conference on Learning Representations},
  series    = {ICLR '26},
  pages     = {42414--42448},
  year      = {2026},
  publisher = {OpenReview.net},
  address   = {Rio de Janeiro, Brazil},
  url       = {https://proceedings.iclr.cc/paper_files/paper/2026/hash/465f22be10e07b301c6ed58f0472f704-Abstract-Conference.html}
}

@inproceedings{katharopoulos2020linear,
  author    = {Katharopoulos, Angelos and Vyas, Apoorv and
               Pappas, Nikolaos and Fleuret, Fran{\c{c}}ois},
  title     = {Transformers are {RNNs}: Fast Autoregressive Transformers
               with Linear Attention},
  booktitle = {Proceedings of the 37th International Conference on Machine Learning},
  series    = {ICML '20},
  volume    = {119},
  pages     = {5156--5165},
  year      = {2020},
  publisher = {PMLR},
  address   = {Virtual Event},
  url       = {https://proceedings.mlr.press/v119/katharopoulos20a.html}
}

@misc{qwen36_35b_a3b,
    title = {{Qwen3.6-35B-A3B}: Agentic Coding Power, Now Open to All},
    url = {https://qwen.ai/blog?id=qwen3.6-35b-a3b},
    author = {{Qwen Team}},
    month = {April},
    year = {2026}
}

@misc{gemmateam2026gemma4,
      title={Gemma 4 Technical Report}, 
      author={Gemma Team},
      year={2026},
      eprint={2607.02770},
      archivePrefix={arXiv},
      primaryClass={cs.CL},
      url={https://arxiv.org/abs/2607.02770}, 
}

@inproceedings{wu2025mcpmark,
  author    = {Zijian Wu and Xiangyan Liu and Xinyuan Zhang and Lingjun Chen and Fanqing Meng and Lingxiao Du and Yiran Zhao and Fanshi Zhang and Yaoqi Ye and Jiawei Wang and Zirui Wang and Jinjie Ni and Yufan Yang and Arvin Xu and Michael Qizhe Shieh},
  title     = {MCPMark: A Benchmark for Stress-Testing Realistic and Comprehensive MCP Use},
  booktitle = {The Fourteenth International Conference on Learning Representations},
  series    = {ICLR '26},
  pages     = {79852--79895},
  year      = {2026},
  publisher = {OpenReview.net},
  address   = {Rio de Janeiro, Brazil},
  url       = {https://proceedings.iclr.cc/paper_files/paper/2026/hash/8138d211ce8790fdfbeeeb9781838a37-Abstract-Conference.html}
}

@inproceedings{vidur,
  author    = {Agrawal, Amey and Kedia, Nitin and Mohan, Jayashree and Panwar, Ashish and
               Kwatra, Nipun and Gulavani, Bhargav S. and Ramjee, Ramachandran and Tumanov, Alexey},
  title     = {{Vidur}: A Large-Scale Simulation Framework for {LLM} Inference},
  booktitle = {Proceedings of Machine Learning and Systems},
  series    = {MLSys '24},
  volume    = {6},
  pages     = {351--366},
  year      = {2024},
  publisher = {mlsys.org},
  address   = {Santa Clara, CA, USA},
  url       = {https://proceedings.mlsys.org/paper_files/paper/2024/hash/b74a8de47d2b3c928360e0a011f48351-Abstract-Conference.html}
}

@misc{MORI,
  title         = {Idleness is Relative: Exploiting Tool-Call Idle Windows for Offloading in Agentic Systems with MORI},
  author        = {Xia, Tian and
                   Xu, Yi and
                   Li, Han-Chen and
                   Li, Zhifei and
                   Chen, Xiaokun and
                   Kang, Hao and
                   Qiao, Yifan and
                   Stoica, Ion},
  year          = {2026},
  eprint        = {2606.00866},
  archivePrefix = {arXiv},
  url           = {https://arxiv.org/abs/2606.00866}
}

@misc{PYTHIA,
      title={Pythia: Exploiting Workflow Predictability for Efficient Agent-Native LLM Serving}, 
      author={Shan Yu and Junyi Shu and Yuanjiang Ni and Kun Qian and Xue Li and Yang Wang and Jinyuan Zhang and Ziyi Xu and Shuo Yang and Lingjun Zhu and Ennan Zhai and Qingda Lu and Jiarong Xing and Youyou Lu and Xin Jin and Xuanzhe Liu and Harry Xu},
      year={2026},
      eprint={2604.25899},
      archivePrefix={arXiv},
      primaryClass={cs.MA},
      url={https://arxiv.org/abs/2604.25899}, 
}

@misc{resum2025,
  author        = {Wu, Xixi and Li, Kuan and Zhao, Yida and Zhang, Liwen and Ou, Litu and
                   Yin, Huifeng and Zhang, Zhongwang and Yu, Xinmiao and Zhang, Dingchu and
                   Jiang, Yong and Xie, Pengjun and Huang, Fei and Cheng, Minhao and
                   Wang, Shuai and Cheng, Hong and Zhou, Jingren},
  title         = {{ReSum}: Unlocking Long-Horizon Search Intelligence via Context
                   Summarization},
  year          = {2025},
  eprint        = {2509.13313},
  archivePrefix = {arXiv},
  url           = {https://arxiv.org/abs/2509.13313}
}

@inproceedings{mem12025,
  author    = {Zhou, Zijian and
               Qu, Ao and
               Wu, Zhaoxuan and
               Kim, Sunghwan and
               Prakash, Alok and
               Rus, Daniela and
               Low, Bryan Kian Hsiang and
               Liang, Paul Pu},
  title     = {{MEM1}: Learning to Synergize Memory and Reasoning for Efficient
                   Long-Horizon Agents},
  booktitle = {The Fourteenth International Conference on Learning Representations},
  series    = {ICLR '26},
  pages     = {58413--58438},
  year      = {2026},
  publisher = {OpenReview.net},
  address   = {Rio de Janeiro, Brazil},
  url       = {https://proceedings.iclr.cc/paper_files/paper/2026/hash/5fc8b3bdfbb9167b5144df5d3fae4616-Abstract-Conference.html}
}

@inproceedings{agentfold2026,
  author    = {Ye, Rui and
               Zhang, Zhongwang and
               Li, Kuan and
               Yin, Huifeng and
               Tao, Zhengwei and
               Zhao, Yida and
               Su, Liangcai and
               Zhang, Liwen and
               Qiao, Zile and
               Wang, Xinyu and
               Xie, Pengjun and
               Huang, Fei and
               Zhou, Jingren and
               Chen, Siheng and
               Jiang, Yong},
  title     = {{AgentFold}: {Long-Horizon} Web Agents with Proactive Context Folding},
  booktitle = {The Fourteenth International Conference on Learning Representations},
  series    = {ICLR '26},
  pages     = {154754--154776},
  year      = {2026},
  publisher = {OpenReview.net},
  address   = {Rio de Janeiro, Brazil},
  url       = {https://proceedings.iclr.cc/paper_files/paper/2026/hash/fa948624dfde013671e72c1a7ca4aebc-Abstract-Conference.html}
}

@misc{compactionrl2026,
  author        = {Li, Yujiang and Hou, Zhenyu and Jing, Yi and Tang, Jie and Dong, Yuxiao},
  title         = {{CompactionRL}: Reinforcement Learning with Context Compaction for
                   Long-Horizon Agents},
  year          = {2026},
  eprint        = {2607.05378},
  archivePrefix = {arXiv},
  url           = {https://arxiv.org/abs/2607.05378}
}

@misc{selfcompact2026,
  author       = {Li, Tianjian and Zhang, Jingyu and Jurayj, William and Wang, Xi and
                   Jin, Chuanyang and Farajtabar, Mehrdad and Nalisnick, Eric and
                   Khashabi, Daniel},
  title        = {Self-Compacting Language Model Agents},
  year         = {2026},
  howpublished = {To appear in \emph{Advances in Neural Information Processing Systems} (NeurIPS '26)},
  url          = {https://neurips.cc/virtual/2026/poster/155897}
}

@misc{memgpt2023,
  author        = {Packer, Charles and Wooders, Sarah and Lin, Kevin and Fang, Vivian and
                   Patil, Shishir G. and Stoica, Ion and Gonzalez, Joseph E.},
  title         = {{MemGPT}: Towards {LLMs} as Operating Systems},
  year          = {2023},
  eprint        = {2310.08560},
  archivePrefix = {arXiv},
  url           = {https://arxiv.org/abs/2310.08560}
}

@inproceedings{snapkv2024,
  author    = {Li, Yuhong and Huang, Yingbing and Yang, Bowen and Venkitesh, Bharat and
               Locatelli, Acyr and Ye, Hanchen and Cai, Tianle and Lewis, Patrick and
               Chen, Deming},
  title     = {{SnapKV}: {LLM} Knows What You are Looking for Before Generation},
  booktitle = {Advances in Neural Information Processing Systems},
  series    = {NeurIPS '24},
  volume    = {37},
  pages     = {22947--22970},
  year      = {2024},
  publisher = {Curran Associates, Inc.},
  address   = {Red Hook, NY, USA},
  doi       = {10.52202/079017-0722},
  url       = {https://proceedings.neurips.cc/paper_files/paper/2024/hash/28ab418242603e0f7323e54185d19bde-Abstract-Conference.html}
}

@misc{onlinekvcompaction2026,
  author        = {Liu, Yujian and Ji, Jiabao and An, Li and Jain, Rohit and
                   Polatkan, Gungor and Zhu, Siyu and Chang, Shiyu},
  title         = {Practical Online {KV} Cache Compaction for {LLM} Agents: An Empirical
                   Study},
  year          = {2026},
  eprint        = {2608.00902},
  archivePrefix = {arXiv},
  url           = {https://arxiv.org/abs/2608.00902}
}

@inproceedings{attnmatching2026,
  author    = {Zweiger, Adam and Fu, Xinghong and Guo, Han and Kim, Yoon},
  title     = {Fast {KV} Compaction via Attention Matching},
  booktitle = {Proceedings of the 43rd International Conference on Machine Learning},
  series    = {ICML '26},
  volume    = {306},
  pages     = {169210--169239},
  year      = {2026},
  publisher = {PMLR},
  address   = {Seoul, South Korea},
  url       = {https://proceedings.mlr.press/v306/zweiger26a.html}
}

@misc{apec2027,
	doi = {10.20944/preprints202611.0780.v1},
	url = {https://doi.org/10.20944/preprints202611.0780.v1},
	year = 2026,
	month = {November},
	publisher = {Preprints},
	author = {Banruo Liu and Haoran Qiu and Íñigo Goiri and Jingran Zhou and Rodrigo Fonseca and Ricardo Bianchini and Fan Lai and Esha Choukse},
	title = {Benchmarking Long-Horizon Agent Context Compaction with CompactionBench},
	journal = {Preprints}
}

@inproceedings{comem2026,
  author    = {Zhang, Yuwei and Dong, Chengyu and Jin, Shuowei and Yu, Changlong and
               Cui, Hejie and Jin, Hongye and Zhang, Xinyang and Bonab, Hamed and
               Lockard, Colin and Chen, Jianshu and Shi, Zhenyu and Shang, Jingbo and
               Li, Xian and Yin, Bing},
  title     = {{CoMem}: Context Management with A Decoupled Long-Context Model},
  booktitle = {International Conference on Machine Learning (ICML)},
  series    = {Proceedings of Machine Learning Research},
  volume    = {306},
  pages     = {155062--155077},
  year      = {2026},
  publisher = {PMLR},
  address   = {Seoul, South Korea},
  url       = {https://proceedings.mlr.press/v306/zhang26ai.html}
}

@misc{parallelcompaction2026,
  author        = {Cim, Musa and Topcu, Burak and Das, Chita and Kandemir, Mahmut Taylan},
  title         = {Parallel Context Compaction for Long-Horizon {LLM} Agent Serving},
  year          = {2026},
  eprint        = {2605.23296},
  archivePrefix = {arXiv},
  url           = {https://arxiv.org/abs/2605.23296}
}

@misc{sleeptime2025,
  author        = {Lin, Kevin and Snell, Charlie and Wang, Yu and Packer, Charles and
                   Wooders, Sarah and Stoica, Ion and Gonzalez, Joseph E.},
  title         = {Sleep-time Compute: Beyond Inference Scaling at Test-time},
  year          = {2025},
  eprint        = {2504.13171},
  archivePrefix = {arXiv},
  url           = {https://arxiv.org/abs/2504.13171}
}

@misc{fastserve2023,
  author        = {Wu, Bingyang and Zhong, Yinmin and Zhang, Zili and Liu, Shengyu and
                   Liu, Fangyue and Sun, Yuanhang and Huang, Gang and Liu, Xuanzhe and
                   Jin, Xin},
  title         = {Fast Distributed Inference Serving for Large Language Models},
  year          = {2023},
  eprint        = {2305.05920},
  archivePrefix = {arXiv},
  url           = {https://arxiv.org/abs/2305.05920}
}

@inproceedings{ltr2024,
  author    = {Fu, Yichao and Zhu, Siqi and Su, Runlong and Qiao, Aurick and
               Stoica, Ion and Zhang, Hao},
  title     = {Efficient {LLM} Scheduling by Learning to Rank},
  booktitle = {Advances in Neural Information Processing Systems 37 (NeurIPS 2024)},
  series    = {NeurIPS '24},
  volume    = {37},
  pages     = {59006--59029},
  year      = {2024},
  publisher = {Curran Associates, Inc.},
  address   = {Red Hook, NY, USA},
  doi       = {10.52202/079017-1882},
  url       = {https://proceedings.neurips.cc/paper_files/paper/2024/hash/6c8985579293e0209bdaa4f21bb1d237-Abstract-Conference.html}
}

@inproceedings{qlm2024,
  author    = {Patke, Archit and Reddy, Dhemath and Jha, Saurabh and Qiu, Haoran and
               Pinto, Christian and Narayanaswami, Chandra and Kalbarczyk, Zbigniew and
               Iyer, Ravishankar},
  title     = {Queue Management for {SLO}-Oriented Large Language Model Serving},
  booktitle = {Proceedings of the ACM Symposium on Cloud Computing},
  series    = {SoCC '24},
  pages     = {18--35},
  year      = {2024},
  publisher = {ACM},
  address   = {New York, NY, USA},
  doi       = {10.1145/3698038.3698523},
  url       = {https://dl.acm.org/doi/10.1145/3698038.3698523}
}

@inproceedings{infercept2024,
  author    = {Abhyankar, Reyna and He, Zijian and Srivatsa, Vikranth and
               Zhang, Hao and Zhang, Yiying},
  title     = {{InferCept}: Efficient Intercept Support for Augmented Large Language
               Model Inference},
  booktitle = {Proceedings of the 41st International Conference on Machine Learning},
  series    = {ICML '24},
  volume    = {235},
  pages     = {81--95},
  year      = {2024},
  publisher = {PMLR},
  address   = {Vienna, Austria},
  url       = {https://proceedings.mlr.press/v235/abhyankar24a.html}
}

@misc{autellix2025,
  author        = {Luo, Michael and Shi, Xiaoxiang and Cai, Colin and Zhang, Tianjun and
                   Wong, Justin and Wang, Yichuan and Wang, Chi and Huang, Yanping and
                   Chen, Zhifeng and Gonzalez, Joseph E. and Stoica, Ion},
  title         = {Autellix: An Efficient Serving Engine for {LLM} Agents as General
                   Programs},
  year          = {2025},
  eprint        = {2502.13965},
  archivePrefix = {arXiv},
  url           = {https://arxiv.org/abs/2502.13965}
}

@misc{tracelab,
  author       = {Zhu, Kan and Jacob, Mathew and Ma, Chenxi and Pan, Yi and
                  Wang, Stephanie and Krishnamurthy, Arvind and Kasikci, Baris},
  title        = {{TraceLab}: Characterizing Coding Agent Workloads for
                  {LLM} Serving},
  year         = {2026},
  eprint       = {2606.30560},
  archivePrefix= {arXiv},
  primaryClass = {cs.DC},
  note         = {Trace release (CC BY 4.0),
                  \url{https://github.com/uw-syfi/TraceLab}, accessed
                  2026-09-11},
  url          = {https://arxiv.org/abs/2606.30560}
}

@misc{tbench21dsh2026,
  author       = {{openguardrails}},
  title        = {{DeepSeek-V4-Flash} on {Terminal-Bench 2.1}: Agent
                  Trajectories},
  year         = {2026},
  publisher    = {Hugging Face},
  note         = {MIT licence, accessed 2026-09-11},
  howpublished = {\url{https://huggingface.co/datasets/openguardrails/terminal-bench-2.1-deepseek-v4-flash-trajectories}}
}

@misc{t1longhorizon2026,
  author       = {{TberiusJunyao}},
  title        = {{T1} Long-Horizon {Terminal-Bench} Trajectories},
  year         = {2026},
  publisher    = {Hugging Face},
  note         = {Accessed 2026-09-11; no longer publicly accessible as of
                  2026-10-27},
  howpublished = {\url{https://huggingface.co/datasets/TberiusJunyao/T1-Long-Horizon-Terminal-Bench-Trajectories}}
}

@misc{exgenticagenttraces2026,
  author       = {{Exgentic}},
  title        = {Agent {LLM} Traces},
  year         = {2026},
  publisher    = {Hugging Face},
  note         = {CDLA-Permissive-2.0, accessed 2026-09-11},
  howpublished = {\url{https://huggingface.co/datasets/Exgentic/agent-llm-traces}}
}

@inproceedings{splitwise2024,
  author    = {Patel, Pratyush and Choukse, Esha and Zhang, Chaojie and
                  Shah, Aashaka and Goiri, {\'I}{\~n}igo and Maleki, Saeed and
                  Bianchini, Ricardo},
  title     = {Splitwise: Efficient Generative {LLM} Inference Using Phase
                  Splitting},
  booktitle = {2024 ACM/IEEE 51st Annual International Symposium on Computer Architecture (ISCA)},
  series    = {ISCA '24},
  pages     = {118--132},
  year      = {2024},
  publisher = {IEEE},
  address   = {Piscataway, NJ, USA},
  doi       = {10.1109/ISCA59077.2024.00019},
  url       = {https://ieeexplore.ieee.org/document/10609649}
}

@misc{azurellmtrace2023,
  author       = {{Microsoft Azure}},
  title        = {Azure {LLM} Inference Trace 2023},
  year         = {2023},
  howpublished = {\url{https://github.com/Azure/AzurePublicDataset/blob/master/AzureLLMInferenceDataset2023.md}},
  note         = {Accessed 2026-09-13}
}

@misc{llmconsistency2026,
  author    = {Abel Yagubyan},
  title        = {How Consistent Are LLM Agents? Measuring Behavioral
                  Reproducibility in Multi-Step Tool-Calling Pipelines},
  year         = {2026},
  eprint       = {2605.28840},
  archivePrefix= {arXiv}
}

@inproceedings{costbench2025,
  author    = {Jiayu Liu and Cheng Qian and Zhaochen Su and Qing Zong and
               Shijue Huang and Bingxiang He and Yi R. (May) Fung},
  title     = {CostBench: Evaluating Multi-Turn Cost-Optimal Planning and
                  Adaptation in Dynamic Environments for LLM Tool-Use Agents},
  booktitle = {Proceedings of the 64th Annual Meeting of the Association for Computational Linguistics (Volume 1: Long Papers)},
  series    = {ACL '26},
  pages     = {12826--12858},
  year      = {2026},
  publisher = {Association for Computational Linguistics},
  address   = {Stroudsburg, PA, USA},
  doi       = {10.18653/v1/2026.acl-long.584},
  url       = {https://aclanthology.org/2026.acl-long.584/}
}

@inproceedings{agentevalsurvey2025,
  author    = {Mahmoud Mohammadi and Yipeng Li and Jane Lo and Wendy Yip},
  title     = {Evaluation and Benchmarking of LLM Agents: A Survey},
  booktitle = {Proceedings of the 31st ACM SIGKDD Conference on Knowledge Discovery and Data Mining V.2},
  series    = {KDD '25},
  pages     = {6129--6139},
  year      = {2025},
  publisher = {ACM},
  address   = {New York, NY, USA},
  doi       = {10.1145/3711896.3736570},
  url       = {https://dl.acm.org/doi/10.1145/3711896.3736570}
}

@inproceedings{atil2024nondeterminism,
  author    = {Atıl, Berk and
               Aykent, Sarp and
               Chittams, Alexa and
               Fu, Liseng and
               Passonneau, Rebecca J. and
               Radcliffe, Evan and
               Rajagopal, Guru Rajan and
               Sloan, Adam and
               Tudrej, Tomasz and
               Ture, Ferhan and
               Wu, Zhe and
               Xu, Lixinyu and
               Baldwin, Breck},
  title     = {Non-Determinism of ``Deterministic'' {LLM} System Settings in Hosted Environments},
  booktitle = {Proceedings of the 5th Workshop on Evaluation and Comparison of NLP Systems},
  series    = {Eval4NLP '25},
  pages     = {135--148},
  year      = {2025},
  publisher = {Association for Computational Linguistics},
  address   = {Stroudsburg, PA, USA},
  doi       = {10.18653/v1/2025.eval4nlp-1.12},
  url       = {https://aclanthology.org/2025.eval4nlp-1.12/}
}

@inproceedings{yuan2025numerical,
  author    = {Jiayi Yuan and Hao Li and Xinheng Ding and Wenya Xie and Yu-Jhe
               Li and Wentian Zhao and Kun Wan and Jing Shi and Xia Hu and
               Zirui Liu},
  title     = {Understanding and Mitigating Numerical Sources of
                   Nondeterminism in {LLM} Inference},
  booktitle = {Advances in Neural Information Processing Systems},
  series    = {NeurIPS '25},
  volume    = {38},
  pages     = {188045--188077},
  year      = {2025},
  publisher = {Curran Associates, Inc.},
  address   = {Red Hook, NY, USA},
  doi       = {10.52202/085713-5653},
  url       = {https://proceedings.neurips.cc/paper_files/paper/2025/hash/f80094a824ba5912d4a2de169c404a40-Abstract-Conference.html}
}

@inproceedings{song2025greedy,
  author    = {Yifan Song and Guoyin Wang and Sujian Li and Bill Yuchen Lin},
  title     = {The Good, The Bad, and The Greedy: Evaluation of {LLMs} Should Not
               Ignore Non-Determinism},
  booktitle = {Proceedings of the 2025 Conference of the North American Chapter of
               the Association for Computational Linguistics (NAACL)},
  pages     = {4195--4206},
  year      = {2025},
  publisher = {Association for Computational Linguistics},
  address   = {Albuquerque, New Mexico},
  doi       = {10.18653/v1/2025.naacl-long.211},
  url       = {https://aclanthology.org/2025.naacl-long.211/}
}

@misc{reliabilitybench2026,
  author    = {Aayush Gupta},
  title         = {ReliabilityBench: Evaluating {LLM} Agent Reliability Under
                   Production-Like Stress Conditions},
  year          = {2026},
  eprint        = {2601.06112},
  archivePrefix = {arXiv}
}

@inproceedings{schick2023toolformer,
  author    = {Schick, Timo and Dwivedi-Yu, Jane and Dess{\`i}, Roberto and
               Raileanu, Roberta and Lomeli, Maria and Hambro, Eric and
               Zettlemoyer, Luke and Cancedda, Nicola and Scialom, Thomas},
  title     = {Toolformer: Language Models Can Teach Themselves to Use Tools},
  booktitle = {Advances in Neural Information Processing Systems 36 (NeurIPS 2023)},
  series    = {NeurIPS '23},
  volume    = {36},
  pages     = {68539--68551},
  year      = {2023},
  publisher = {Curran Associates, Inc.},
  address   = {Red Hook, NY, USA},
  doi       = {10.52202/075280-2997},
  url       = {https://proceedings.neurips.cc/paper_files/paper/2023/hash/d842425e4bf79ba039352da0f658a906-Abstract-Conference.html}
}

@misc{claudecode,
  title        = {Claude Code},
  author       = {{Anthropic}},
  year         = {2025},
  howpublished = {\url{https://www.anthropic.com/claude-code}},
  note         = {Accessed October 2026}
}

@misc{openaicodex,
  title        = {Codex {CLI}: A Lightweight Coding Agent for the Terminal},
  author       = {{OpenAI}},
  year         = {2025},
  howpublished = {\url{https://github.com/openai/codex}},
  note         = {Accessed October 2026}
}

@misc{qwen3,
  title         = {Qwen3 Technical Report},
  author        = {{Qwen Team}},
  year          = {2025},
  eprint        = {2505.09388},
  archivePrefix = {arXiv}
}

@misc{deepseekv3,
  title         = {{DeepSeek-V3} Technical Report},
  author        = {{DeepSeek-AI}},
  year          = {2024},
  eprint        = {2412.19437},
  archivePrefix = {arXiv}
}

@misc{belcak2026longcontext,
  title         = {You Don't Need Long Context},
  author        = {Belcak, Peter and Muralidharan, Saurav and Jiang, Jindong and Fu, Yonggan and Byeon, Wonmin and Ye, Hanrong and Angilly, Ryan and Taghibakhshi, Ali and Liu, Shih-Yang and Ouyang, Siru and Heinrich, Greg and Molchanov, Pavlo},
  year          = {2026},
  eprint        = {2611.05306},
  archivePrefix = {arXiv},
  primaryClass  = {cs.AI}
}

@misc{glm45,
  title         = {{GLM-4.5}: Agentic, Reasoning, and Coding ({ARC}) Foundation Models},
  author        = {{GLM-4.5 Team}},
  year          = {2025},
  eprint        = {2508.06471},
  archivePrefix = {arXiv},
  primaryClass  = {cs.CL},
  url           = {https://arxiv.org/abs/2508.06471}
}

@inproceedings{lee2024infinigen,
  author    = {Wonbeom Lee and Jungi Lee and Junghwan Seo and Jaewoong Sim},
  title     = {{InfiniGen}: Efficient Generative Inference of Large Language
               Models with Dynamic {KV} Cache Management},
  booktitle = {18th USENIX Symposium on Operating Systems Design and Implementation (OSDI 24)},
  series    = {OSDI '24},
  pages     = {155--172},
  year      = {2024},
  publisher = {USENIX Association},
  address   = {Berkeley, CA, USA},
  url       = {https://www.usenix.org/conference/osdi24/presentation/lee}
}

@inproceedings{sun2025shadowkv,
  author    = {Hanshi Sun and Li-Wen Chang and Wenlei Bao and Size Zheng and
               Ningxin Zheng and Xin Liu and Harry Dong and Yuejie Chi and
               Beidi Chen},
  title     = {{ShadowKV}: {KV} Cache in Shadows for High-Throughput
               Long-Context {LLM} Inference},
  booktitle = {Proceedings of the 42nd International Conference on Machine Learning},
  series    = {ICML '25},
  volume    = {267},
  pages     = {57355--57373},
  year      = {2025},
  publisher = {PMLR},
  address   = {Vancouver, Canada},
  url       = {https://proceedings.mlr.press/v267/sun25b.html}
}

@misc{li2026longhorizonterminalbench,
  title         = {Long-Horizon-Terminal-Bench: Testing the Limits of Agents on
                   Long-Horizon Terminal Tasks with Dense Reward-Based Grading},
  author        = {Zongxia Li and Zhongzhi Li and Yucheng Shi and Ruhan Wang and
                   Junyao Yang and Zhichao Liu and Xiyang Wu and Anhao Li and
                   Chengsong Huang and Ergan Shang and Yue Yu and Ninghao Liu and
                   Lichao Sun and Haotao Mi and Leowei Liang},
  year          = {2026},
  eprint        = {2607.08964},
  archivePrefix = {arXiv},
  primaryClass  = {cs.AI}
}

@misc{deepseekai2026deepseekv4,
  title         = {{DeepSeek-V4}: Towards Highly Efficient Million-Token Context
                   Intelligence},
  author        = {{DeepSeek-AI}},
  year          = {2026},
  eprint        = {2606.19348},
  archivePrefix = {arXiv},
  primaryClass  = {cs.CL}
}

@misc{qwen3.5,
  title  = {{Qwen3.5}: Towards Native Multimodal Agents},
  author = {{Qwen Team}},
  month  = {February},
  year   = {2026},
  howpublished = {\url{https://qwen.ai/blog?id=qwen3.5}}
}

@misc{agentservesim,
  title         = {{AgentServeSim}: Serving-System Simulation and Policy Search for {LLM} Agent Programs},
  author        = {Rajib, Rakibul Hasan and Zheng, Mengxin and Lou, Qian},
  year          = {2026},
  eprint        = {2606.09613},
  archiveprefix = {arXiv},
  primaryclass  = {cs.CL}
}

@inproceedings{vidur-agent,
  author    = {Kim, Donghwan and Singh, Prakhar and Min, Younghoon and Kim, Jongryool and Park, Jongse and Maeng, Kiwan},
  title     = {Characterizing How Complex Agentic {AI} Systems Handle General Tasks: A Trace-Based Simulation Study},
  booktitle = {2026 IEEE International Symposium on Workload Characterization (IISWC)},
  series    = {IISWC '26},
  pages     = {72--86},
  year      = {2026},
  publisher = {IEEE},
  address   = {Piscataway, NJ, USA},
  doi       = {10.1109/IISWC71020.2026.00018},
  url       = {https://ieeexplore.ieee.org/document/11766179/}
}

@misc{frontier,
  title         = {Frontier: Towards Comprehensive and Accurate {LLM} Inference Simulation},
  author        = {Feng, Yicheng and Tan, Xin and Deng, Yangtao and Jiang, Yimin and Zhu, Yibo and Xu, Hong},
  year          = {2026},
  eprint        = {2605.21312},
  archiveprefix = {arXiv},
  primaryclass  = {cs.DC}
}

@misc{glm5,
  title         = {{GLM-5}: from Vibe Coding to Agentic Engineering},
  author        = {{GLM-5 Team}},
  year          = {2026},
  eprint        = {2602.15763},
  archiveprefix = {arXiv},
  primaryclass  = {cs.LG}
}

@misc{kimik3,
  title        = {{Kimi K3}: Open Frontier Intelligence},
  author       = {{Kimi Team}},
  year         = {2026},
  howpublished = {\url{https://github.com/MoonshotAI/Kimi-K3}},
  note         = {Technical report}
}

@misc{aiconfigurator,
  title         = {{AIConfigurator}: Lightning-Fast Configuration Optimization for Multi-Framework {LLM} Serving},
  author        = {Xu, Tianhao and
                   Liu, Yiming and
                   Lu, Xianglong and
                   Zhao, Yijia and
                   Zhou, Xuting and
                   Feng, Aichen and
                   Chen, Yiyi and
                   Shen, Yi and
                   Zhou, Qin and
                   Chen, Xumeng and
                   Sherstyuk, Ilya and
                   Li, Haorui and
                   Thakkar, Rishi and
                   Hamm, Ben and
                   Li, Yuanzhe and
                   Huang, Xue and
                   Wu, Wenpeng and
                   Shanbhag, Anish and
                   Kim, Harry and
                   Chen, Chuan and
                   Lai, Junjie},
  year          = {2026},
  eprint        = {2601.06288},
  archiveprefix = {arXiv},
  primaryclass  = {cs.DC}
}

@misc{tbscience,
  title        = {{Terminal-Bench Science}: Benchmarking {AI} Agents on Scientific Workflows},
  author       = {{The Terminal-Bench Team}},
  year         = {2026},
  howpublished = {\url{https://www.tbench.ai/news/terminal-bench-science-0-1}},
  note         = {Version 0.1}
}

@inproceedings{cacheblend,
author = {Yao, Jiayi and Li, Hanchen and Liu, Yuhan and Ray, Siddhant and Cheng, Yihua and Zhang, Qizheng and Du, Kuntai and Lu, Shan and Jiang, Junchen},
title = {CacheBlend: Fast Large Language Model Serving for RAG with Cached Knowledge Fusion},
year = {2025},
isbn = {9798400711961},
publisher = {Association for Computing Machinery},
address = {New York, NY, USA},
url = {https://doi.org/10.1145/3689031.3696098},
doi = {10.1145/3689031.3696098},
booktitle = {Proceedings of the Twentieth European Conference on Computer Systems},
pages = {94--109},
numpages = {16},
location = {Rotterdam, Netherlands},
series = {EuroSys '25}
}
